\documentclass[a4paper,USenglish,cleveref,autoref,pdflatex]{lipics-v2021}

\graphicspath{{./images/}}

\title{Maniposynth:\\Bimodal Tangible Functional Programming}
\titlerunning{Bimodal Tangible Functional Programming} 

\author{Brian Hempel}{University of Chicago, USA \and \url{http://brianhempel.com/}}{brian@brianhempel.com} {} {}

\author{Ravi Chugh}{University of Chicago, USA \and \url{http://people.cs.uchicago.edu/~rchugh/}}{rchugh@cs.uchicago.edu} {} {}

\authorrunning{B. Hempel and R. Chugh} 

\Copyright{Brian Hempel and Ravi Chugh} 



\nolinenumbers 

\EventEditors{Karim Ali and Jan Vitek}
\EventNoEds{2}
\EventLongTitle{36th European Conference on Object-Oriented Programming (ECOOP 2022)}
\EventShortTitle{ECOOP 2022}
\EventAcronym{ECOOP}
\EventYear{2022}
\EventDate{June 6--10, 2022}
\EventLocation{Berlin, Germany}
\EventLogo{}
\SeriesVolume{222}
\ArticleNo{16}

\usepackage[utf8]{inputenc}
\usepackage{relsize}
\usepackage{pbsi}

\usepackage{newunicodechar}
\newunicodechar{🎉}{\includegraphics[height=0.8em]{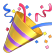}}

\usepackage[dvipsnames]{xcolor} 

\begin{document}

\newcommand{\ms}{\ensuremath{\textsc{Maniposynth}}}
\newcommand{\msFancy}{{\bsifamily (The Magnificent) Maniposynth}}
\newcommand{\ie}{\textit{i.e.}}
\newcommand{\eg}{\textit{e.g.}}

\newcommand{\vcentered}[1]{\begin{tabular}{l}#1\end{tabular}}

\newcommand{\valigntop}[1]{\vtop{\null\hbox{#1}}}

\renewcommand{\subsectionautorefname}{Section}

\maketitle

\begin{figure}[hbt]
  \frame{\includegraphics[width=\textwidth]{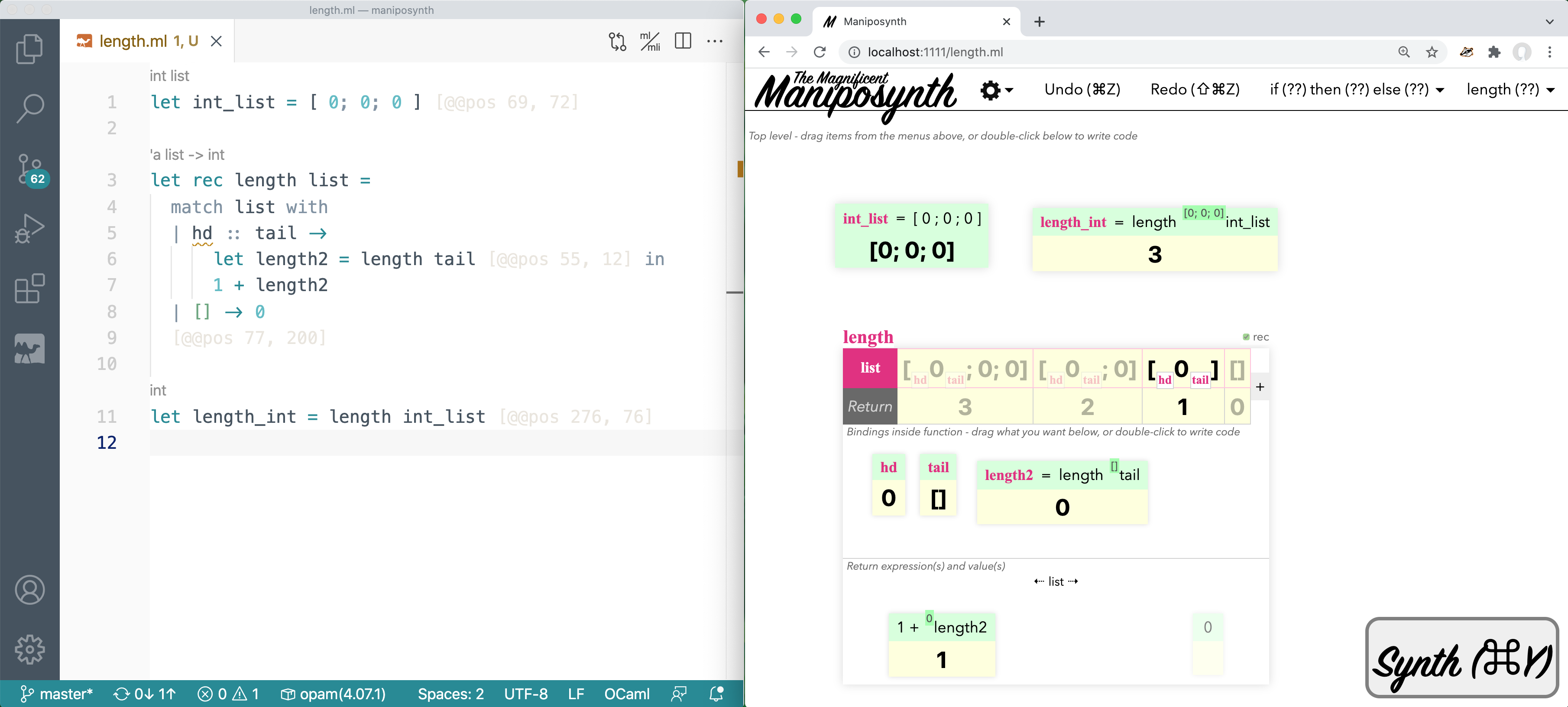}}  
  \caption{A list \texttt{length} function implemented in \ms{}.}
  \label{fig:teaser}
\end{figure}

\begin{abstract}
Traditionally, writing code is a non-graphical, abstract, and linear process. Not everyone is comfortable with this way of thinking at all times. Can programming be transformed into a graphical, concrete, non-linear activity?
While nodes-and-wires~\cite{TheOnLineGraphicalSpecificationOfComputerProcedures} and blocks-based~\cite{LogoBlocks} programming environments do leverage graphical direct manipulation, users perform their manipulations on abstract syntax tree elements, which are still abstract. Is it possible to be more concrete—could users instead directly manipulate live program values to create their program?

We present a system, \ms{}, that reimagines functional programming as a non-linear workflow where program expressions are spread on a 2D canvas. The live results of those expressions are continuously displayed and available for direct manipulation. The non-linear canvas liberates users to work out-of-order, and the live values can be interacted with via drag-and-drop. Incomplete programs are gracefully handled via hole expressions, which allow \ms{} to offer program synthesis. Throughout the workflow, the program is valid OCaml code which the user may inspect and edit in their preferred text editor at any time.

With \ms{}'s direct manipulation features, we created 38 programs drawn from a functional data structures course. We additionally hired two professional OCaml developers to implement a subset of these programs. We report on these experiences and discuss to what degree \ms{} meets its goals of providing a non-linear, concrete, graphical programming workflow.

\end{abstract}

\section{Introduction}
\label{sec:introduction}

Graphical, direct manipulation interfaces~\cite{DirectManipulation} are the paradigm most users are familiar with when they operate computers. Graphical interfaces are powerful and vastly extend the reach of computing. Nevertheless, the most powerful computer activity---programming---has proven resistant to manifestation in a graphical, direct manipulation form. Most programming is primarily a text-only activity. Can general-purpose programming be reimagined in a graphical, direct manipulation interface? Experts might find productivity gains and novices might find a more approachable environment to accomplish their goals.

Most existing graphical programming approaches present the abstract syntax tree (AST) elements as items to be manipulated with the cursor. Nodes-and-wires programming environments~\cite{TheOnLineGraphicalSpecificationOfComputerProcedures}, such as LabVIEW~\cite{LabVIEW}, present program expressions as boxes whose inputs and outputs are connected by wires. Blocks-based environments, such as Scratch~\cite{Scratch}, present the program expressions as puzzle pieces that snap together. And structure editors, such as Barista~\cite{Barista}, allow certain manipulations of program expressions as structured entities rather than as a naive string of text. All these approaches center the AST as the object of interaction.
Even more concrete than AST elements, however, are the \emph{values} a program produces during execution. Humans are concrete thinkers before we are abstract thinkers, and teachers know that the best way to explain is through examples. So, is there a way to write programs via direct manipulation on \emph{values} instead of on AST elements?

The Eros environment~\cite{Eros} demonstrated a compelling answer to this question. Eros reimagined the programming space not as a program in text (as in traditional coding) nor as a draftsman's drawing of operations connected by wires (as in nodes-and-wires programming), but as a 2D canvas of malleable values. These \emph{tangible values (TVs)} were primarily partially applied functions, rendered with (graphically editable) example arguments for their unapplied inputs, with the corresponding example output displayed below. The user could select the output of one TV, the input of another TV (of corresponding type), and compose the two together into a new TV.
%

Eros highlighted that \emph{non-linear editing} and pure functional programming are complementary. Without side effects, the order of computation is negligible. The user may gather the parts they need in any order and worry later about how to assemble them.
Alas, the standard practice of writing functional programs as linear, textual code obscures this opportunity for non-linearity. Placing values on a 2D canvas instead highlights it.

Non-linearity matters. Not all humans are linear thinkers, and not even all programmers think linearly at all times. (How often are large blocks of code written top-to-bottom from scratch?) A non-linear environment can offer a creative space more inviting to folks whose standard workflow naturally entails concrete exploration rather than abstract planning.

While Eros highlighted how non-linear editing dovetails with pure functional programming, its mechanism for composing TVs may tip the balance too far from the abstract in favor of the concrete. Once a value has been composed, it obscures \emph{how} it came to be. TVs are labeled with a brief expression, but this one line is inadequate for any computation of modest size. Moreover, once composed, how does one change the computation that produced a TV? Value manipulation alone may be inadequate for carefully specifying abstract algorithms. Perhaps there is a middle ground that allows both non-linear, concrete direct manipulation on values \emph{and} traditional editing of ordinary code.
That middle ground is the subject of this paper. In particular, we seek to answer the question:

\begin{quote}
\emph{How can the approachability of non-linear direct manipulation on concrete values be melded with the time-proven flexibility of text-based coding?}
\end{quote}

\subparagraph*{Design Goals}

We aim to create a programming interface with the following properties:

\begin{alphaenumerate}
	\item \textbf{Value-Centric.} Like Eros, and unlike most visual programming environments, we want values—not AST elements—to be centered in the display and, as much as possible, be the object of the user's direct manipulations.
	\item \textbf{Non-Linear.} To support non-linear thinkers and exploratory programming, we want to allow the user to gather the parts they need out of order, and compose them later.
	\item \textbf{Synthesis.} How to integrate recent advances in program synthesis into a practical workflow remains an open question. A value-centric interface is a natural environment to specify assertions on those displayed values and to fulfill those assertions with a synthesizer—we want to explore this.
	\item  \textbf{Bimodal.}
		Ideally, a visual programming environment would not sacrifice the unique affordances of textual code—its concision and its amenability to an ecosystem of existing tooling (such as text editors, language servers, and version control). We want to offer a bimodal interface that simultaneously offers a non-linear graphical editing interface \emph{alongside} a text-editable, traditional representation of the program's code.
\end{alphaenumerate}

\subparagraph*{Contributions}

To show how value-centric non-linear editing can meld with traditional text-based programming, we implemented a value-centric, non-linear, bimodal programming environment with synthesis features called \msFancy{}. We demonstrate both how non-linear visual editing can integrate with linear code, as well as show novel editing features made possible by the value-centric display.

To gain an initial understanding of the system, we implemented an external corpus of 38 example programs.
For additional insights, we conducted an in-depth exploratory study with two external professional functional programmers, whose feedback informed the evolution of \ms{}. We describe their use of the tool and discuss additional observations through investigative lenses from the Cognitive Dimensions of Notation framework~\cite{CognitiveDimensions}.

Section \ref{sec:overview} introduces \ms{} with a running example. Section \ref{sec:implementation} describes the technical implementation of the tool and the synthesizer. \autoref{sec:evaluation} presents insights from implementing a corpus of examples and the qualitative user study. \autoref{sec:relatedWork} presents related work, and \autoref{sec:last} discusses avenues for continued exploration.

\section{Overview Example}
\label{sec:overview}

To provide an overview of \ms{}, we follow a fictional programmer named Baklava as she re-implements the list \verb+length+ function from scratch. \autoref{fig:teaser} shows the final result. A video of this example, as well as an artifact to follow along, are available online~\cite{ManiposynthDotOrg}.

\ms{} is a locally running web application designed to be opened in a web browser alongside the user's preferred text editor. Baklava creates a blank text file named \verb+length.ml+ on her computer, starts \ms{} in that directory, and navigates to \url{http://localhost:1111/length.ml} in her web browser. She positions her browser window side-by-side with Visual Studio Code~\cite{VSCode} and is ready to begin.

\subsection{List Length, Without Synthesis}

\begin{figure}[bt]
\centering

\begin{minipage}{.25\textwidth}%
  \includegraphics[height=0.8in]{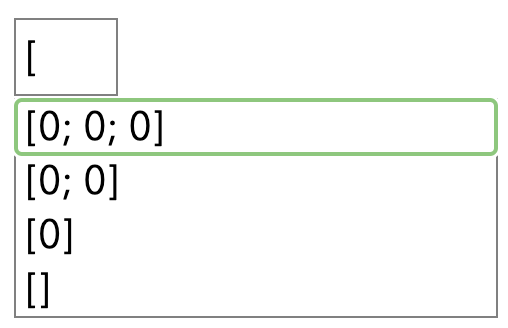}
  \caption{List literals in autocomplete menu.}
  \label{fig:autocompleteListLiterals}

\end{minipage}%
\hfill
\begin{minipage}{.72\textwidth}%

  \includegraphics[height=0.8in]{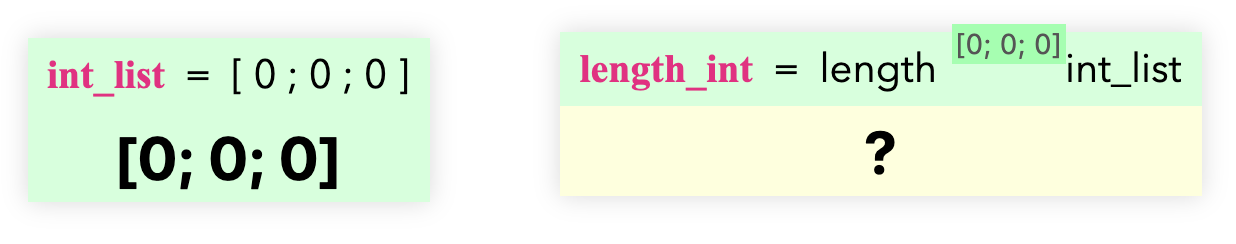}
  \caption{Tangible values (TVs) for the example list binding and the example call to \texttt{length}.}
  \label{fig:exampleTVs}

\end{minipage}

\end{figure}

To start, \ms{} displays a blank white 2D canvas. Because \ms{} is a live programming environment, Baklava starts by creating an example list so she can see the \verb+length+ operation on concrete data. Double-clicking on the canvas opens up a text box to add new code; Baklava does so and types an open bracket \verb+[+. Because writing example data is common in \ms{}, concrete literals up to a fixed size are offered as autocomplete options (auto-generated from the data constructors in scope, \autoref{fig:autocompleteListLiterals}). Baklava selects the list literal \verb+[0; 0; 0]+ from the autocomplete options and hits Enter.

In the code, a new let-binding for the list is inserted at the top-level of \verb+length.ml+ and automatically given the name \verb+int_list+. On the canvas, this binding is represented as a box displaying (in clockwise order, \autoref{fig:exampleTVs}, left) the binding pattern (\verb+int_list+), the binding expression (\verb+[ 0 ; 0 ; 0 ]+), and the result value below (also \verb+[ 0; 0; 0 ]+, but bigger). These three elements together in a box form a \emph{tangible value} in \ms{}. The box may be repositioned on the 2D canvas, and the coordinates of the position are stored in the code as an AST attribute annotation on the binding, written \verb+[@@pos 152, 49]+ in the code. Arbitrary attribute annotations are supported by the standard OCaml AST which allow these properties to be preserved across program transformations. Baklava has installed a VS Code plugin to dim these attributes in the code to avoid becoming distracted by them.

To begin work on the \verb+length+ function, Baklava now creates an example call to the function: on the canvas, she double-clicks to add new code and types \verb+length int_list+. As before, a new binding is inserted in the code (named \verb+length_int+) and an associated tangible value (TV) appears on the canvas (\autoref{fig:exampleTVs}, right).

The \verb+length_int+ TV has two differences from the previous \verb+int_list+ TV. First, its result value (displayed as \verb+?+, explained below) has a yellow background—this indicates the result is \emph{not} simply a constant introduced in the expression: it came from computation elsewhere. Second, the \verb+int_list+ variable usage in the TV's expression bears a superscript indicating the value of \verb+int_list+, namely \verb+[0; 0; 0]+.

\begin{figure}[bt]
  \centering
  \includegraphics[height=1.9in]{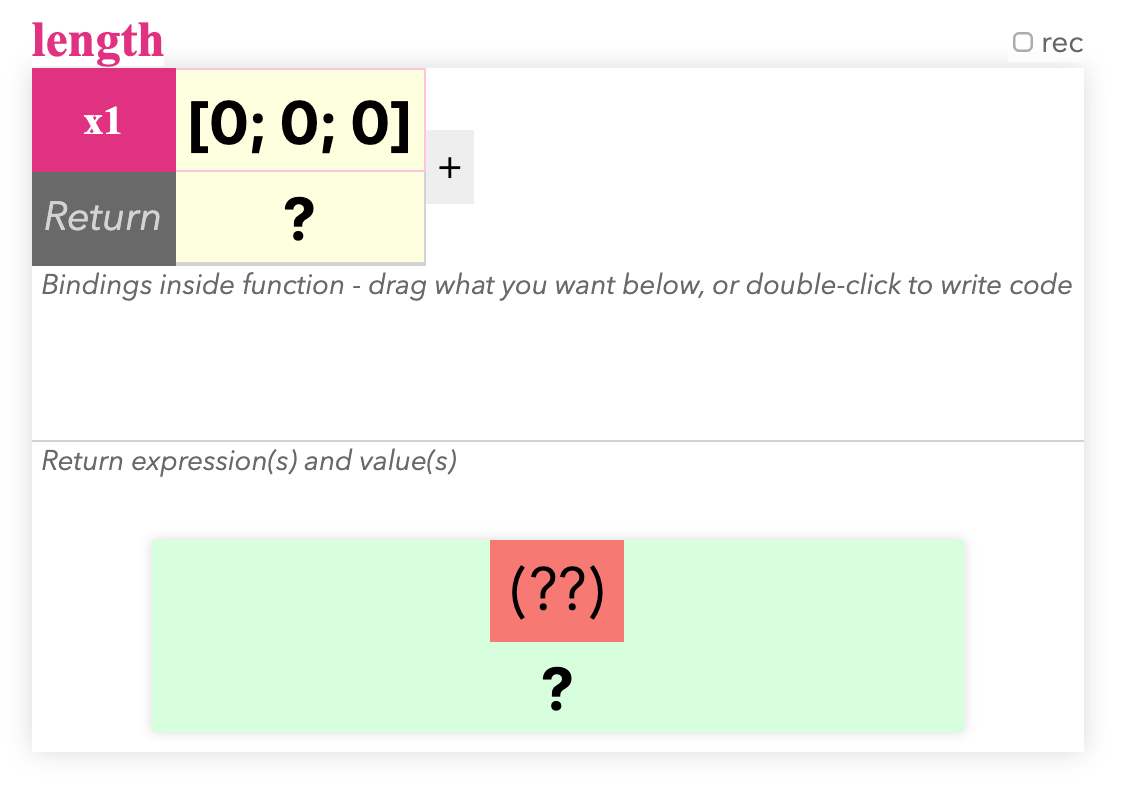}
  \caption{Tangible value for the function skeleton binding \texttt{let length x1 = (??)}.}
  \label{fig:functionSkeletonTV}
\end{figure}

In \ms{}, using an undefined variable—in this case, \verb+length+—automatically inserts a new let-binding (TV) for that variable. Because Baklava used \verb+length+ as a function, a new function skeleton was inserted in the code (\verb+let length x1 = (??)+).

Function TVs are displayed specially on the canvas (\autoref{fig:functionSkeletonTV}). Immediately below the function name, a \emph{function IO grid} displays the function input and output values encountered during execution. Immediately below the IO grid is a blank white area which is a \emph{subcanvas} for the bindings (TVs) inside the function, of which there are none yet. Below the subcanvas is a (non-movable) TV for the return expression and overall result value of the function. Currently the function return expression is a \emph{hole expression}, written \verb+(??)+. Hole expressions are placeholders, expected to be filled in later. For this reason, they are displayed larger than normal expressions (to make them easier targets for clicking) and have a slowly pulsing red background (to remind the user that the program is unfinished). While the \verb+(??)+ syntax is supported by OCaml's editor tooling (Merlin~\cite{Merlin} and its language server protocol wrapper~\cite{OCamlLSP}), programs with holes are ordinarily not executable. To continue to provide live feedback in the presence of holes, \ms{} evaluates hole expressions \verb+(??)+ to a \emph{hole value}, displayed as \verb+?+. This hole value \verb+?+ is the current return value of the function shown below \verb+(??)+—in green because it was introduced by the immediate expression above—and also shown in the ``Return'' row of the IO grid as well as, back on the main top-level canvas, in the result value of the \verb+length int_list+ function call.

\begin{figure}[bt]
  \setlength{\tabcolsep}{0.2em}
  \centering
  \begin{tabular}{rcrc}
  \vcentered{(a)}
  & \vcentered{\includegraphics[height=1.73in]{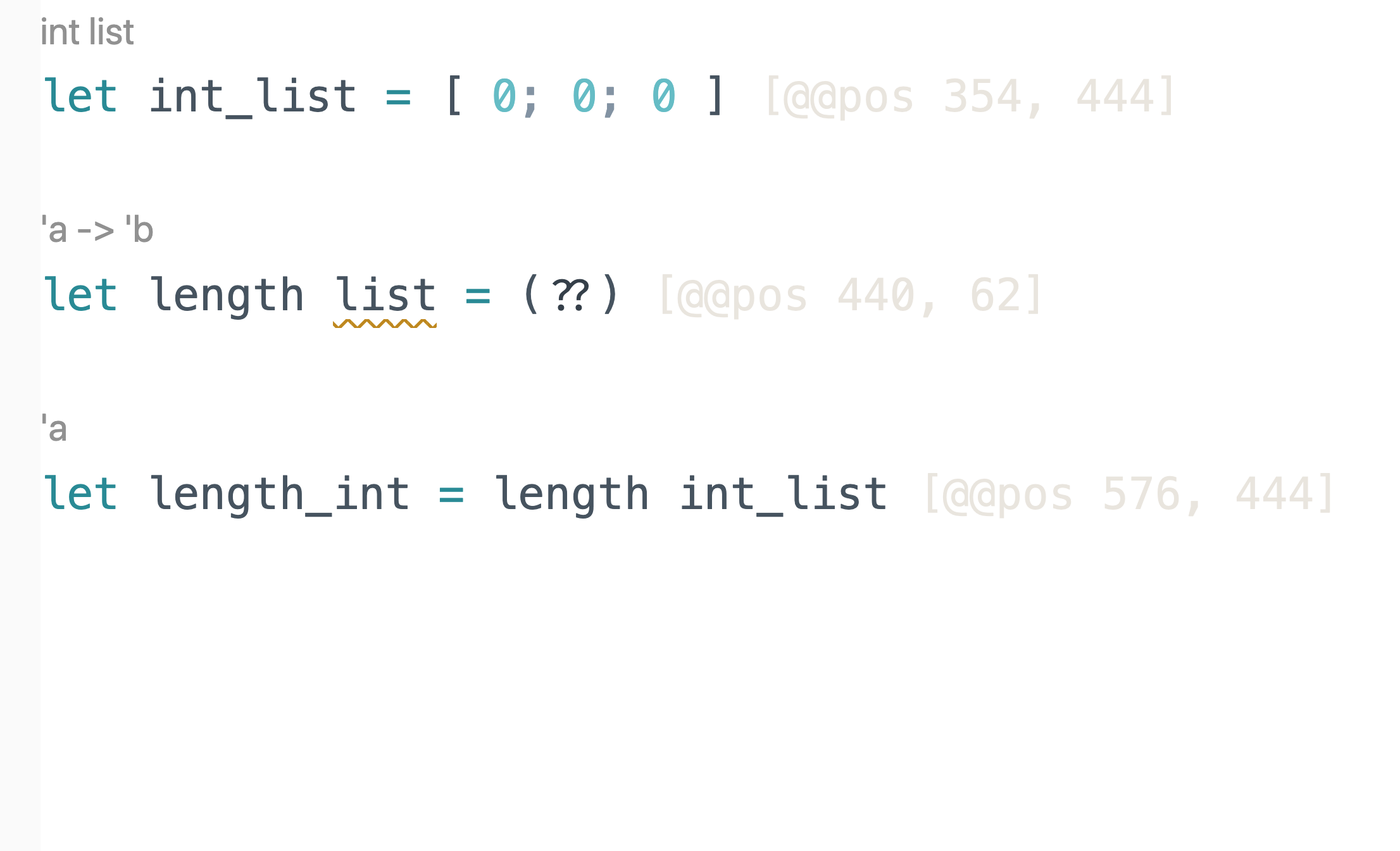}}
  & \vcentered{(b)}
  & \vcentered{\includegraphics[height=1.73in]{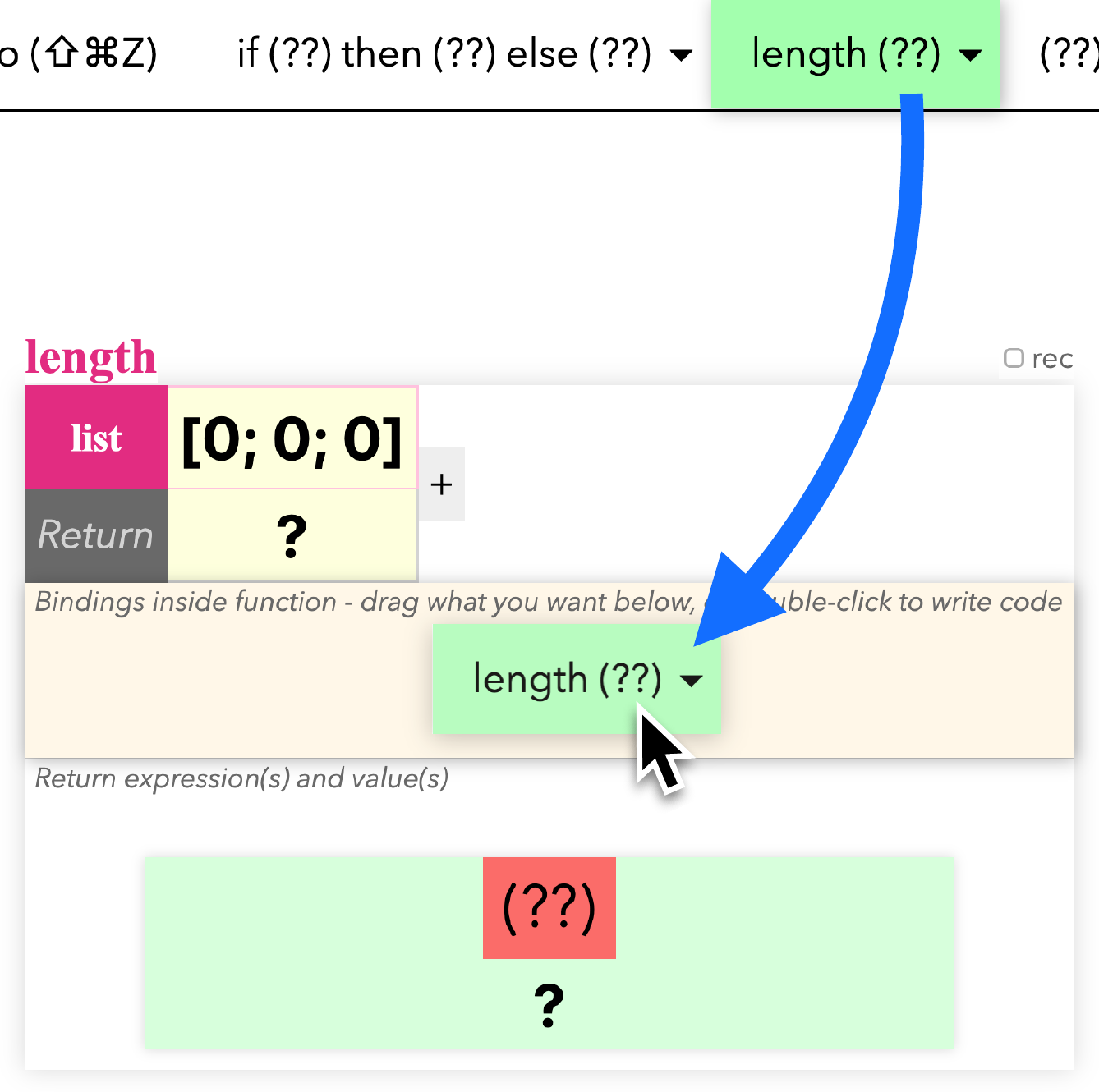}} \\
  \vcentered{(c)}
  & \vcentered{\includegraphics[height=1.73in]{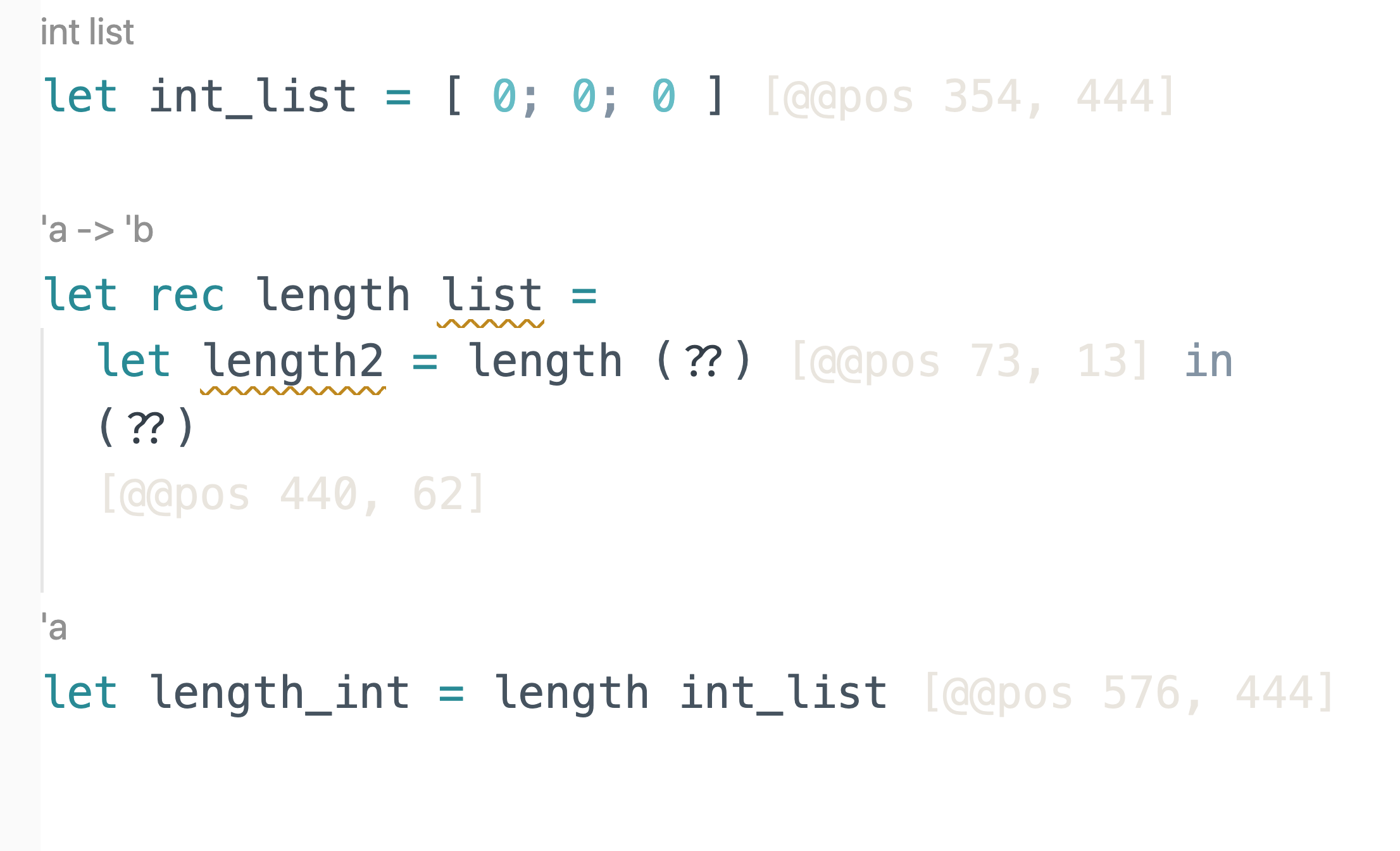}}
  & \vcentered{(d)}
  & \vcentered{\includegraphics[height=1.73in]{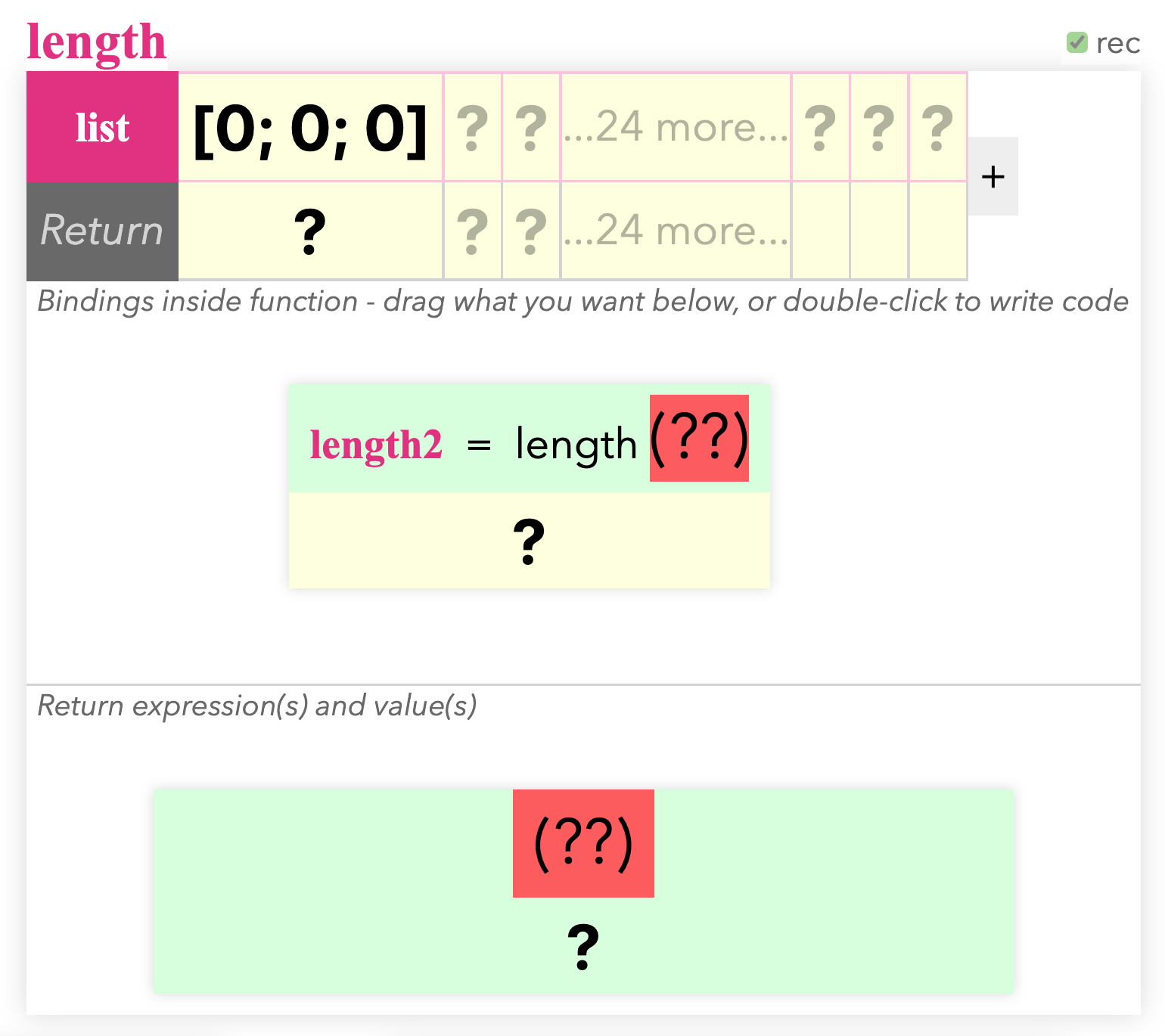}}
\end{tabular}
  \caption[Creating a recursive call]{Creating a recursive call. (a) Code before. (b) Dragging a new \texttt{length} call into the \texttt{length} function. (c) Resulting code and (d) resulting \texttt{length} function TV.}
  \label{fig:dragLengthIntoItself}
\end{figure}

Baklava does not like the default \verb+x1+ parameter name in the \verb+length+ function and wants to rename it. Most items in \ms{} can be double-clicked to perform a text edit. Baklava double-clicks the pink-background \verb+x1+ to rename the variable (\underline{p}atterns are \underline{p}ink), and writes the name \verb+list+ instead. \autoref{fig:dragLengthIntoItself}a shows the code at this point.

\begin{figure}[bt]
  \centering
  \setlength{\tabcolsep}{0.2em}
  \begin{tabular}{cc}
  \valigntop{(a)} & \valigntop{\includegraphics[width=5.3in]{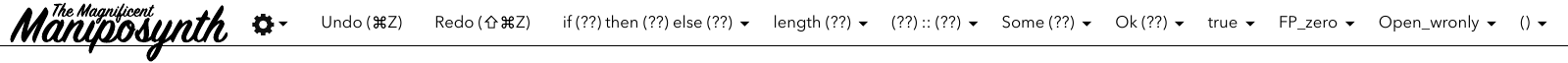}}
  \end{tabular}
  \begin{tabular}{cccccc}
  \raisebox{-4pt}[0pt][0pt]{\valigntop{(b)}} & \valigntop{\includegraphics[height=1.79in]{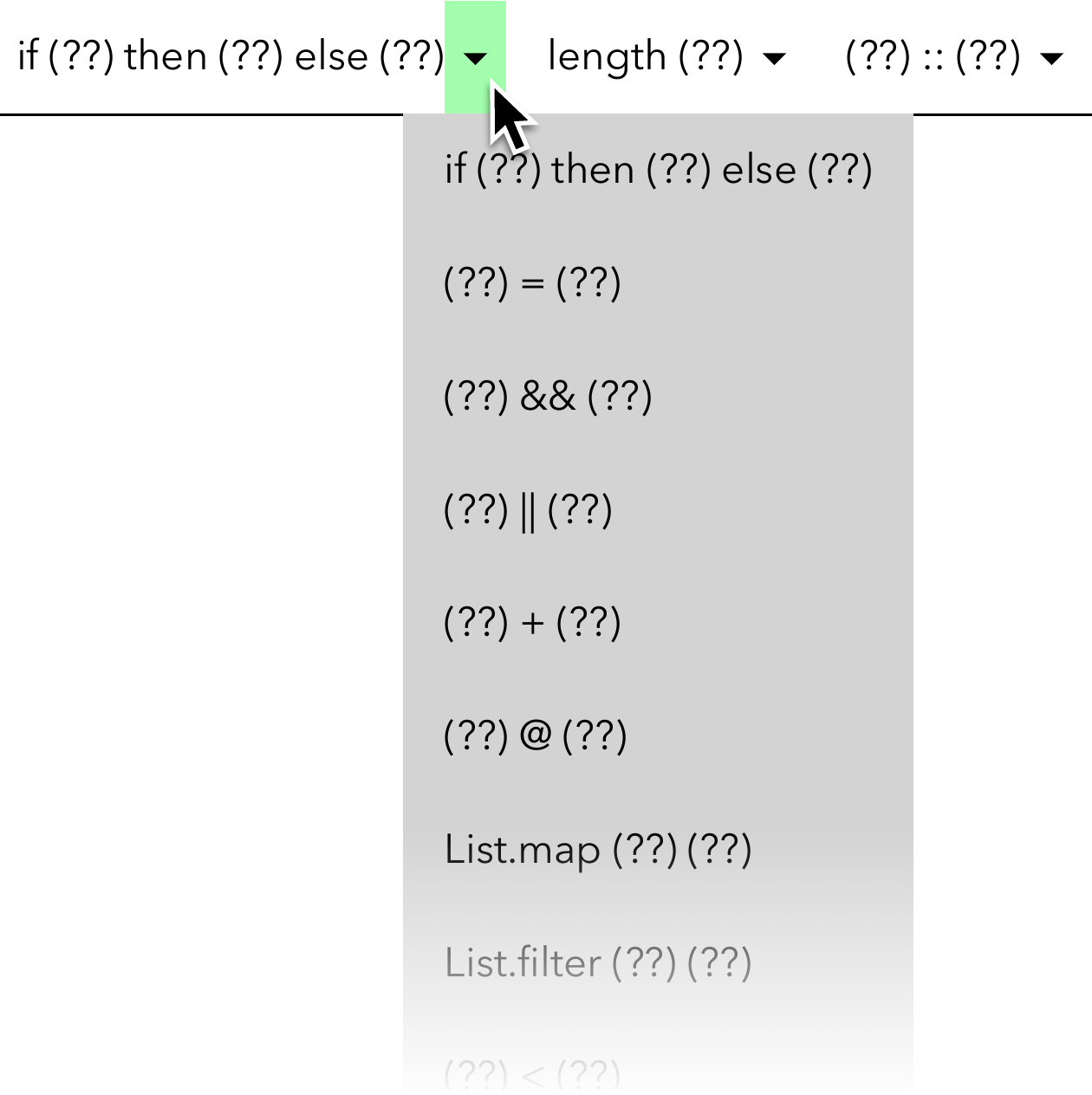}} & \hspace{4em}\raisebox{-4pt}[0pt][0pt]{\valigntop{(c)}} & \valigntop{\includegraphics[height=0.37in]{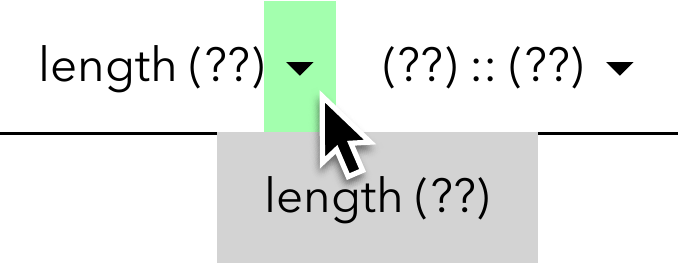}} & \hspace{4em}\raisebox{-4pt}[0pt][0pt]{\valigntop{(d)}} & \valigntop{\includegraphics[height=1.1in]{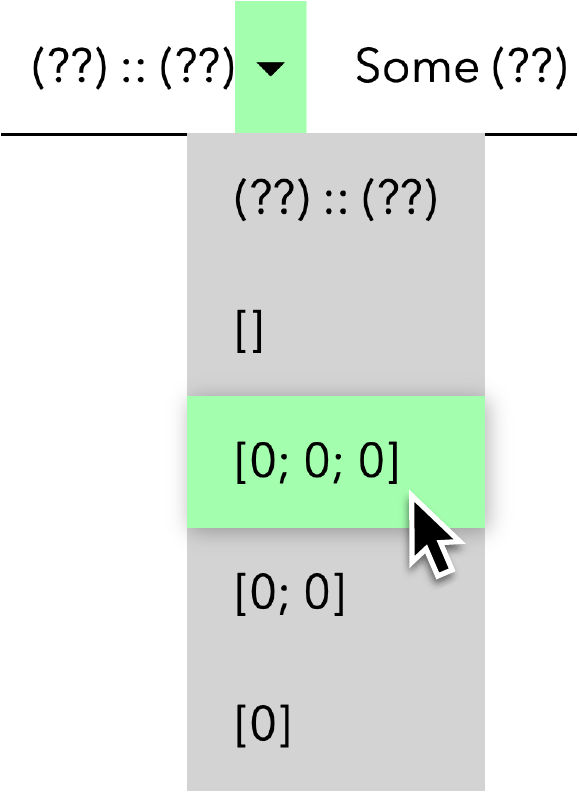}}
 \end{tabular}

  \caption[Toolbar]{(a) The toolbar, including menus for (b) skeleton expressions, (c) functions defined in the current file, and expressions auto-generated from data types in scope, \eg{} the lists shown in (d).}
  \label{fig:toolbar}
\end{figure}

A goal of \ms{} is to allow non-linear editing—the ability to make incremental progress toward a solution. Baklava knows she must make a recursive call to the \verb+length+ function, so, without thinking hard about what might come after, she decides to add \verb+length (??)+ inside \verb+length+. She could double-click and type this code, but typing \verb+(??)+ requires some finger gymnastics.
\ms{} supports a large number of drag-and-drop interactions. Any green expression can be dragged to a new location to duplicate that expression: dropping on an existing expression (\eg{} a hole) replaces the existing expression, while dropping on a (sub)canvas inserts a new binding (TV). Values and patterns can also be dragged to expressions or (sub)canvases—when hovering over a value or pattern, a tooltip shows what expression will be inserted. Finally, a \emph{toolbar} at the top of the window (\autoref{fig:toolbar}a) offers menus containing skeleton expressions: the first menu offers common expressions such as \verb+if (??) then (??) else (??)+ (\autoref{fig:toolbar}b); the second menu offers functions defined in this file (\autoref{fig:toolbar}c); and the remaining menus offer constructors and automatically generated example values of the types in scope (\eg{} \autoref{fig:toolbar}d; the expressions are the same as those offered by autocomplete). User-defined custom data types, if any, also appear as menus.

Baklava drags \verb+length (??)+ from the toolbar into the subcanvas for her \verb+length+ function (\autoref{fig:dragLengthIntoItself}b). A \verb+length2 = length (??)+ binding is created in the code (\autoref{fig:dragLengthIntoItself}c) and an associated TV appears inside \verb+length+ (\autoref{fig:dragLengthIntoItself}d). \ms{} also changes the top-level \verb+let length = ...+ into \verb+let rec length = ...+.

Because \verb+(??)+ produces hole value \verb+?+ instead of crashing, the \verb+length+ function is now diverging as \verb+length (??)+ calls \verb+length (??)+ which calls \verb+length (??)+ and so on. \ms{} uses fueled execution to cut off infinite loops and keep functioning. In the function IO grid, extra columns show these calls (\autoref{fig:dragLengthIntoItself}d), but other than understanding why these extra columns are there, Baklava need not mind that her program is momentarily divergent.

\begin{figure}[bt]
  \setlength{\tabcolsep}{0.2em}
  \centering
  \begin{tabular}{cc}
  \vcentered{(a)}
  & \vcentered{\includegraphics[height=0.7in]{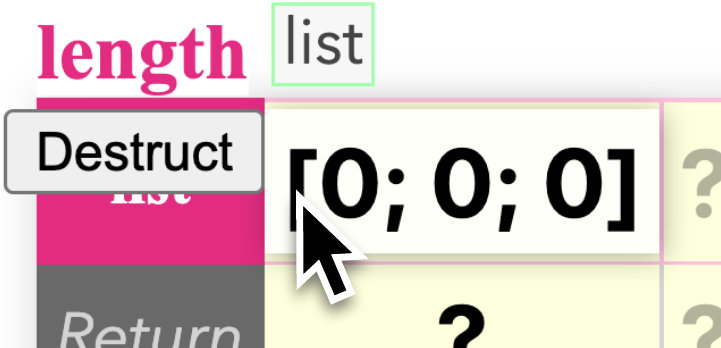}}
  \\
  \\
  \end{tabular} \\
  \begin{tabular}{clc}
  \vcentered{(b)}
  & \vcentered{\includegraphics[height=1.60in]{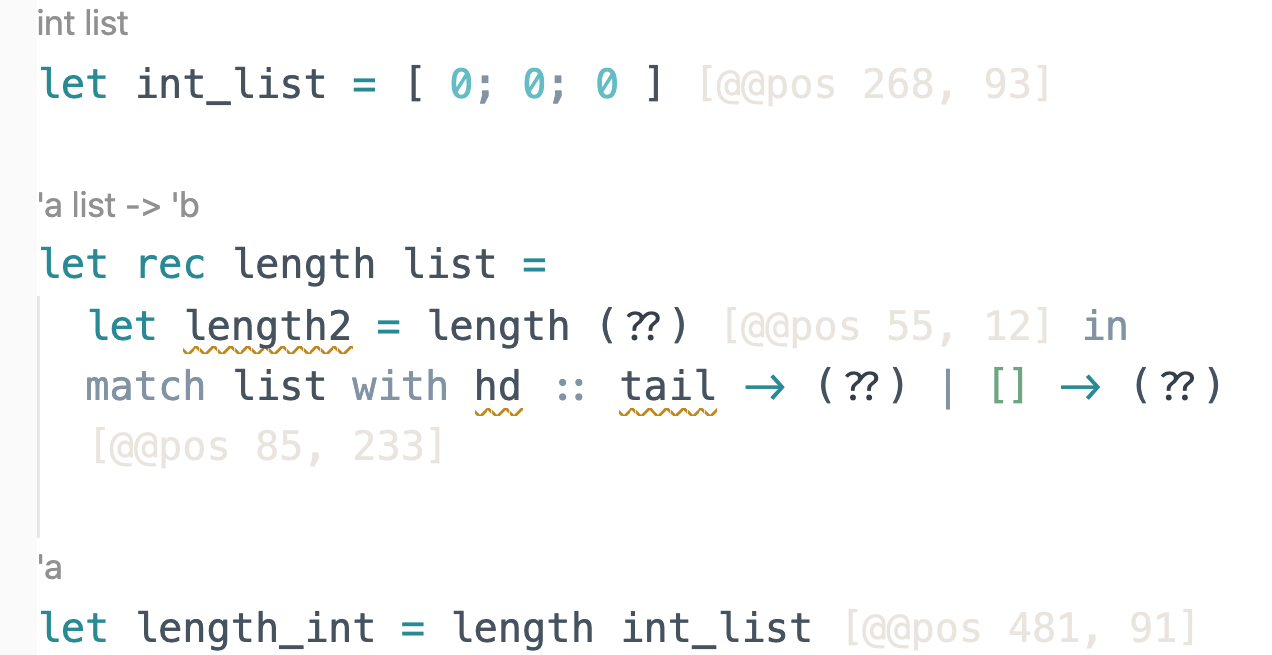}}
  & \vcentered{\includegraphics[height=1.70in]{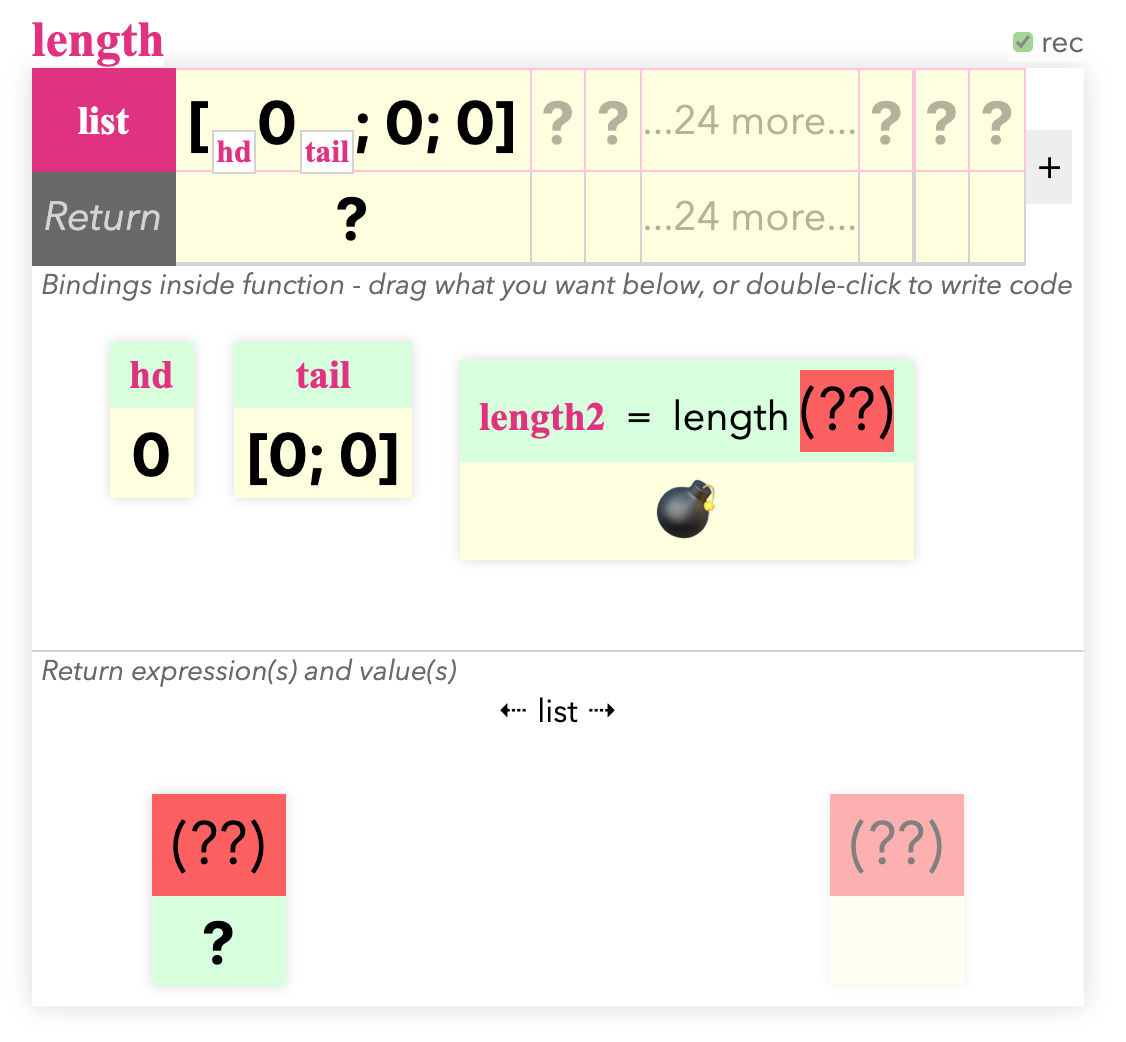}} \\
  \\
  \vcentered{(c)}
  & \vcentered{\includegraphics[height=0.93in]{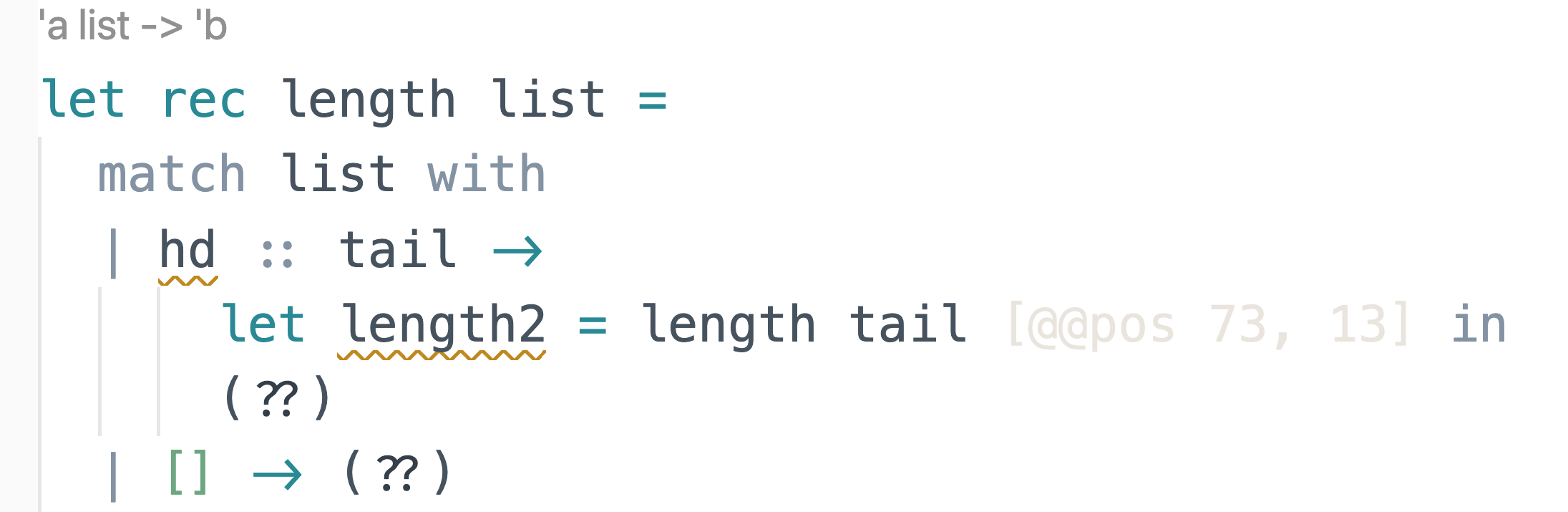}}
  & \vcentered{\includegraphics[height=0.93in]{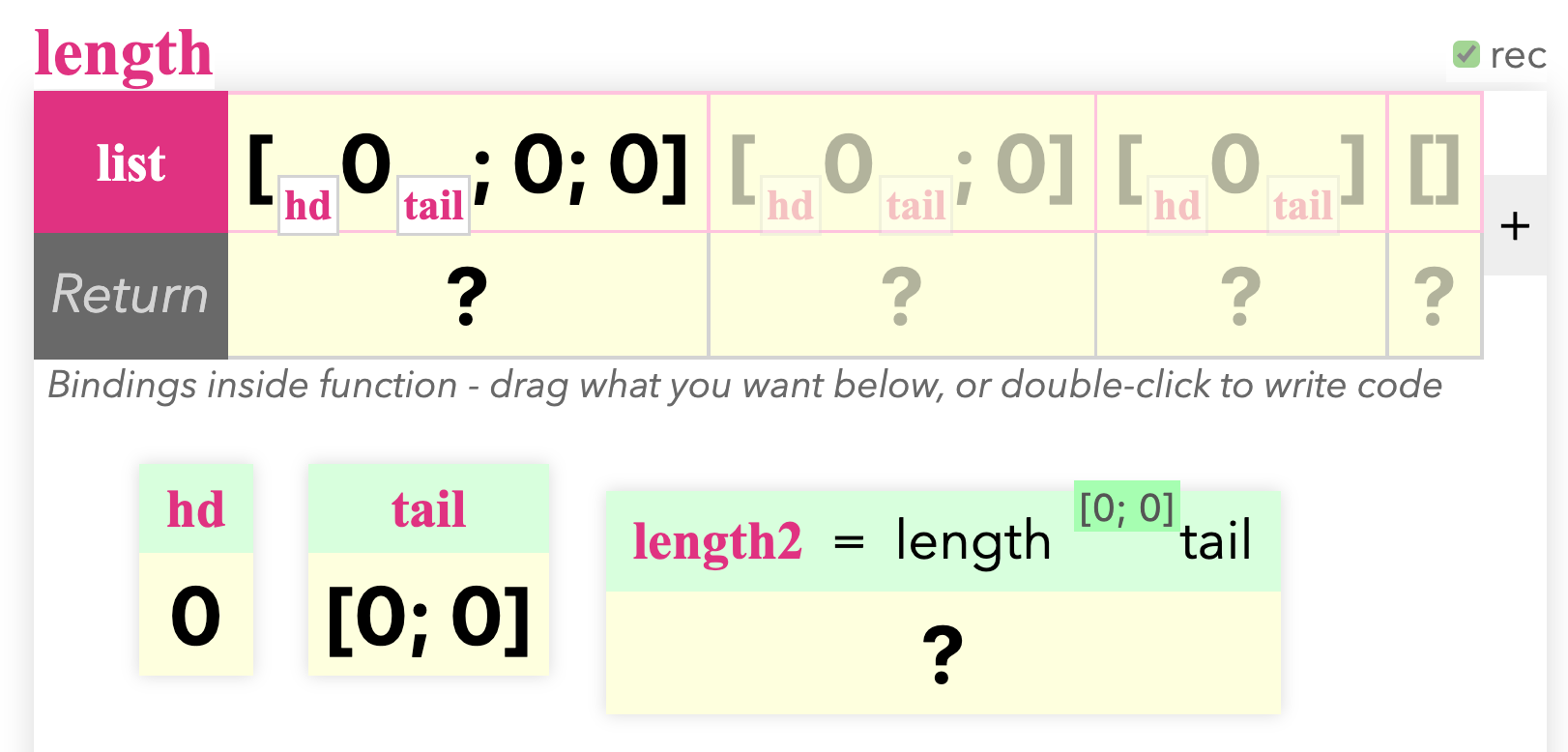}}
  \end{tabular}

  \caption[Destructing]{(a)~A~``Destruct'' button appears when the cursor hovers over an input that is an ADT value. (b)~Code and TV for \texttt{length} function after destructing the \texttt{[0; 0; 0]} input value. (c)~After dragging the tail value \texttt{[0; 0]} to the red hole argument \texttt{(??)} for the recursive \texttt{length} call. }
  \label{fig:destructingEtc}
\end{figure}

Baklava wants the recursive call to operate on the tail of the input list. When she moves the cursor over the input list in the IO grid, a ``Destruct'' button appears (\autoref{fig:destructingEtc}a), which she clicks. As shown in \autoref{fig:destructingEtc}b, a \verb+match+ statement (\ie{} case split) is added, with holes for the return expression of each branch. On the display, there are a number of visual changes. In the IO grid, \verb+hd+ and \verb+tail+ pink subscripts appear inside the input list \verb+[0; 0; 0]+, labeling the subvalues that are now bound to names by the \verb+match+ statement. To make these bindings even clearer, they are also represented as two new TVs in the function subcanvas. Finally, the function now has two possible return expressions: both appear as (non-movable) TVs at the bottom of the function, one is grayed out indicating it is not the branch taken when the input is \verb+[0; 0; 0]+ (the column currently selected in the IO grid). Above the two return TVs is an indication of the scrutinee, ``← list →'', which allows editing of the scrutinee expression.

Now that the list tail is exposed on the subcanvas, Baklava drags it (either the pink \verb+tail+ name or the \verb+[0; 0]+ value below it) onto the hole in \verb+length (??)+, transforming it into \verb+length tail+. In her code, the binding is moved from the top-level of the function into the branch in which \verb+tail+ exists (\autoref{fig:destructingEtc}c). Because \ms{} embraces non-linear editing, the user should not have to worry about binding order—bindings will automatically be shuffled around as necessary to place items in the appropriate scope.

The additional calls from the recursion appear in the function IO grid, each still returning hole value \verb+?+ (\autoref{fig:destructingEtc}c, right). Baklava would like to edit the base case, so she looks for the column in the IO grid where the input is \verb+[]+, and then clicks that  column to bring that \emph{call frame} into focus. Call frames are effectively equivalent to runtime stack frames. The TVs not executed on that call are grayed out (\verb+hd+, \verb+tail+, \verb+length2+, and the return for the \verb+hd::tail+ branch). Baklava double-clicks the no longer grayed-out return expression \verb+(??)+ for the base case and sets it to the constant \verb+0+. (She could also have double-clicked the green-background hole value \verb+?+; values are rendered with a green background when double-clicking them will effect an edit on the expression immediately above.)

\begin{figure}[bt]
  \centering
  \includegraphics[height=3.5in]{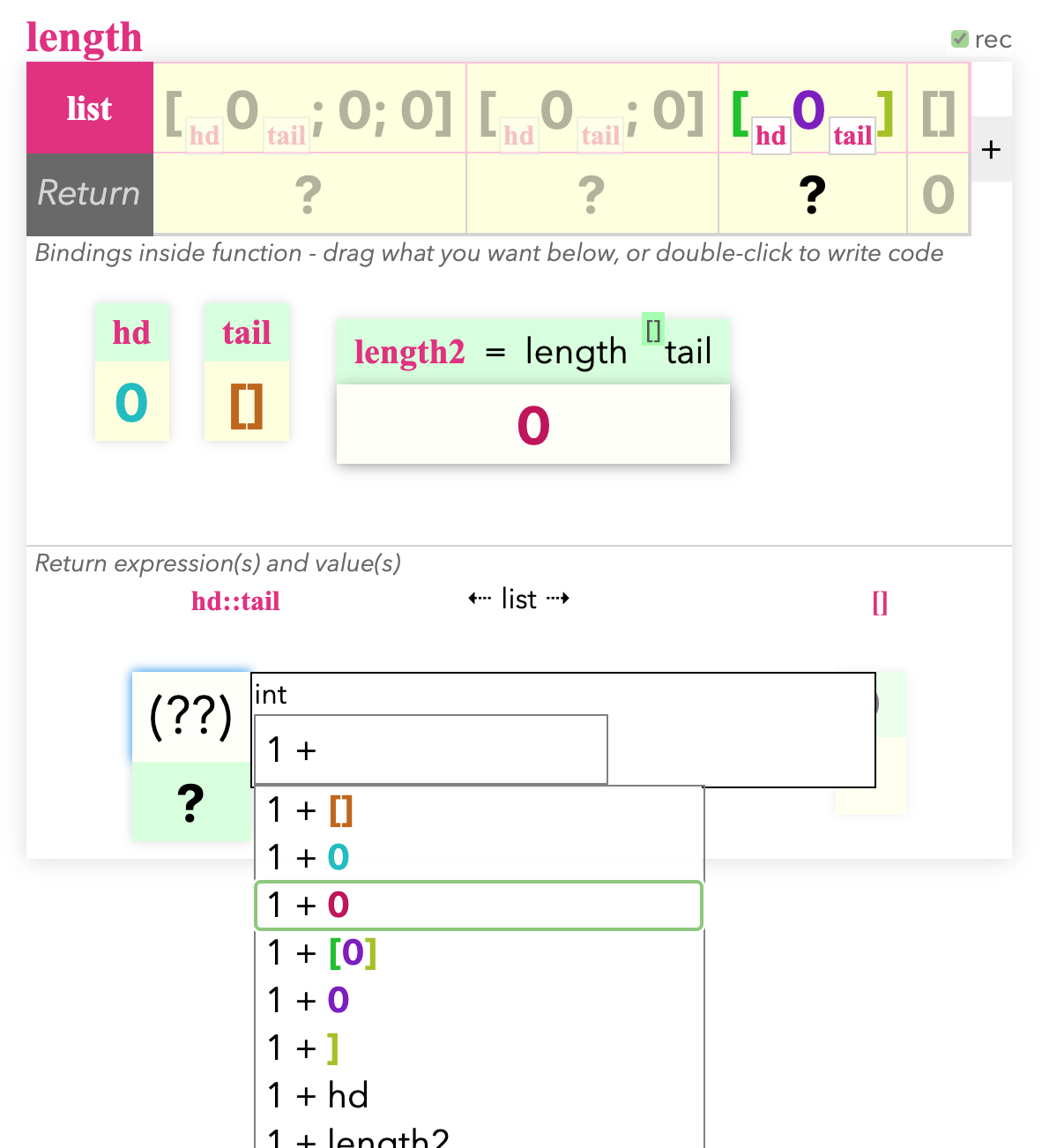}
  \caption{Autocompleting to a value in scope.}
  \label{fig:autocompleteToValue}
\end{figure}

Baklava now clicks the second-to-last call frame in the IO grid to bring into focus the call where the input is \verb+[0]+. The return expression for this branch is still \verb+(??)+. She notes that the TV for the \verb+length tail+ call now displays a result value of 0. Baklava double-clicks the return expression \verb+(??)+ and, after typing ``\verb-1 + -'' she pauses (\autoref{fig:autocompleteToValue}). When she began to type, \ms{} recolored the displayed values in scope using different colors, and now, looking at the autocomplete options, she sees \texttt{1 + \textcolor[rgb]{0.039,0.737,0.761}{\textbf{0}}}, \texttt{1 + \textcolor[rgb]{0.761,0.039,0.353}{\textbf{0}}}, and \texttt{1 + \textcolor[rgb]{0.494,0.039,0.761}{\textbf{0}}} among the possible autocompletions—each with a different color \texttt{\textbf{0}} corresponding to a similarly colored \texttt{\textbf{0}} value elsewhere on screen. The maroon \textcolor[rgb]{0.761,0.039,0.353}{\texttt{\textbf{0}}} is the return from \verb+length tail+, so she chooses that. The branch return expression becomes \verb-1 + length2-, and Baklava can now see in the IO grid that her function returns the correct value for all inputs (\autoref{fig:teaser}).

\subsection{Undo and Delete}

\ms{} supports undo/redo. Additionally, any expression may be selected by a single click and then transformed to a hole by pressing the Delete key. Entire let-binding TVs can similarly be selected and deleted, removing them from the program. Uses of the binding must be deleted before deleting the binding itself—otherwise \ms{} will immediately recreate a binding to satisfy the unbound variable uses.

\subsection{Value-Centric Shortcuts, and Synthesis}

There are usually multiple ways to complete a task in \ms{}. Below are a few variations Baklava might have performed instead.

\subparagraph*{Drag-to-Extract}
To extract the list tail for use in a recursive call to \verb+length+, Baklava clicked ``Destruct'' on the input value and dragged the resulting \verb+tail+ name to the \verb+length (??)+ call. The explicit ``Destruct'' step can be skipped. Because \ms{} aims is to make values live as much as possible, \emph{subvalues} can also be manipulated. Baklava could have hovered over the portion of the input list \verb+[0; 0; 0]+ that is the tail of that list, namely \verb+; 0; 0]+, and dragged that subvalue directly to her \verb+length (??)+ call without pressing ``Destruct''. The destruction will be performed automatically, producing the same code.

\subparagraph*{Autocomplete-to-Extract}
Similarly, visible subvalues are also offered as autocompletions. Perhaps the fastest way to create list \verb+length+ is, immediately after the \verb+length+ function skeleton is created, to double-click the return hole expression, type ``\verb-1 + length -'', and then finish the new expression by selecting \texttt{\textcolor[rgb]{0.761,0.039,0.353}{\textbf{;~0;~0]}}}, the tail of the input list, from the autocomplete options. The expression \verb-1 + length tail- and the needed pattern match will be inserted, leaving only the base case to fill in.

\begin{figure}[t]
  \centering
  \begin{tabular}{cc}
  \setlength{\tabcolsep}{0pt}
  \hspace{-0.20in}
  \vcentered{\includegraphics[width=1.05in]{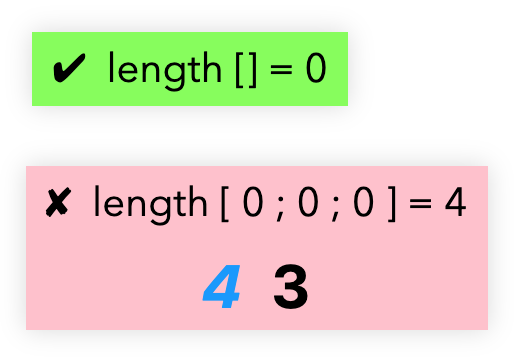}} &
  \hspace{-0.15in}
  \vcentered{\includegraphics[width=4.40in]{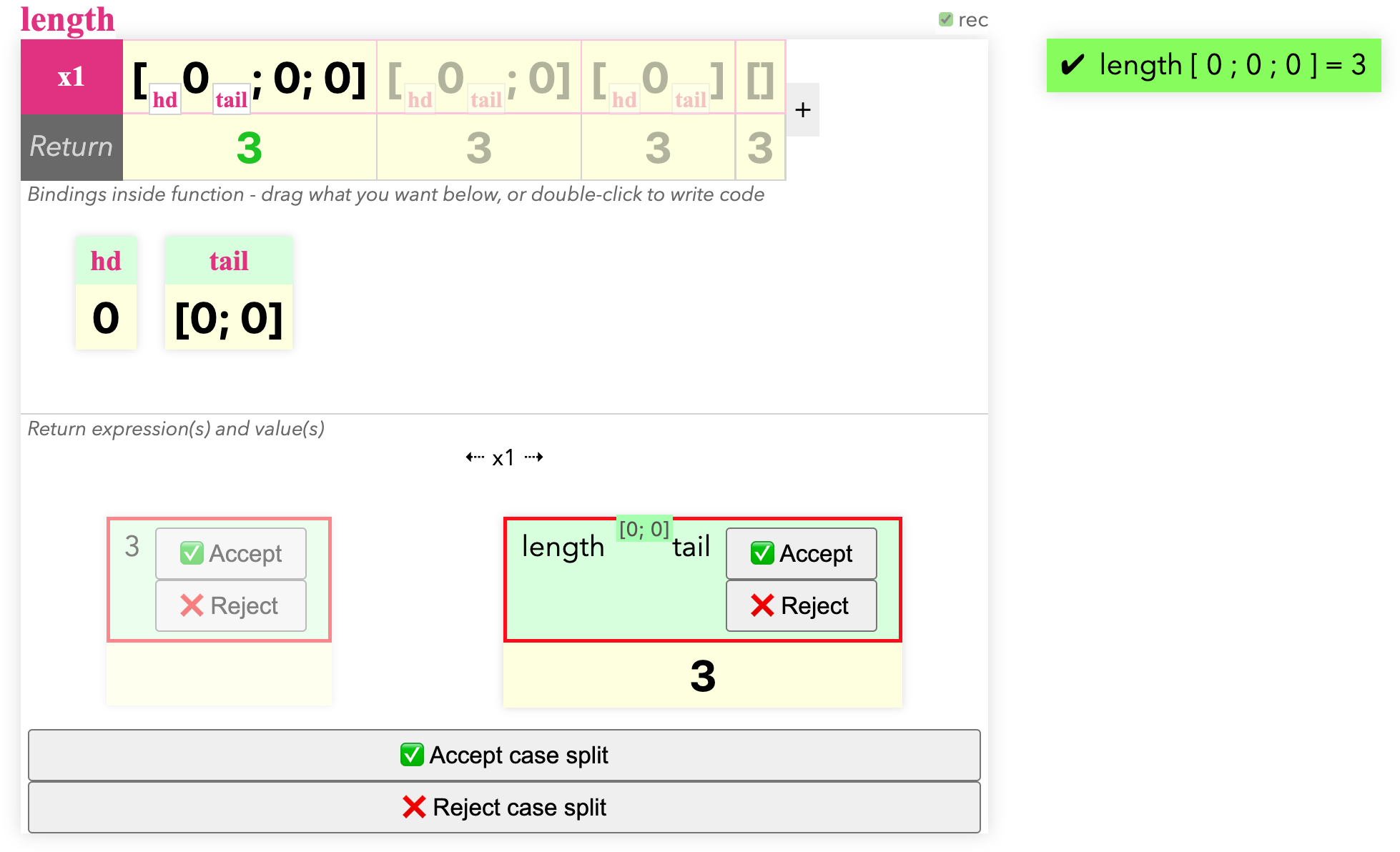}} \\
  \hspace{-0.20in}
  (a) &
  \hspace{-0.15in}
  (b)
  \end{tabular}
  \caption[Synthesis result display]{(a) A satisfied and unsatisfied assertion. (b) An undesired synthesis result. After rejecting the return expressions (or the entire case split), the next result will be correct.}
  \label{fig:assertsAndIncorrectSynthResult}
\end{figure}

\subparagraph*{Assertions}
Baklava started with an example call to \verb+length+. To remind herself of the goal, she could have created an assertion instead: typing \verb+length [0; 0; 0] = 3+ on the top-level canvas will add an \verb+assert+ statement instead of a named binding. Assertions are rendered in red when unsatisfied, and both the expected result (in blue) and the actual result (in black) are shown. When satisfied, an assertion turns green and its result is hidden (\autoref{fig:assertsAndIncorrectSynthResult}a).
Assertions can also be added via the function IO grid: clicking the ``+'' button at the right of the IO grid creates a new column in the grid, allowing Baklava to fill in input values and expected output. Upon hitting enter, the column is reified by adding a new assertion at the top-level, so that the function is indeed called with the specified arguments.

\subparagraph*{Program Synthesis}
Assertions facilitate \emph{programming by example (PBE)}~\cite{pbeSurvey}, a workflow currently available in Microsoft Excel~\cite{FlashFill} but not yet in ordinary programming settings. After asserting \verb+length [0; 0; 0] = 3+, Baklava might have clicked the ``Synth'' button in the lower-right corner of the UI. \ms{} will use type-and-example based synthesis (inspired by \textsc{Myth}~\cite{Myth}) to guess hole fillings until the assertion is satisfied or the synthesizer gives up (after between 10 and 40 seconds). The synthesizer incorporates a simple statistics model and other heuristics to improve result quality (\autoref{sec:synthesizer}). In this scenario, with only the single assertion, \ms{} instantly finds a filling that creates the proper case split, but places \verb+3+ in the base case and \verb+length tail+ in the recursive case (\autoref{fig:assertsAndIncorrectSynthResult}b). The result is incorporated into the code, but presented to Baklava with buttons prompting her to ``Accept'' or ``Reject'' parts of the synthesized expression. Rejected expressions are transformed back into holes, with an annotation telling the synthesizer to avoid that expression in the future. Baklava accepts the case split, but rejects its two return expressions. When she clicks ``Synth'' again, the desired return expressions are discovered she accepts them.

\section{Implementation}
\label{sec:implementation}

The main features of \ms{} now demonstrated, next we describe how it works.

\subsection{Architecture Overview}

\ms{} is a web application written in $\sim$8600 lines of OCaml (excluding the interpreter) and $\sim$2000 lines of Javascript.
OCaml's compiler tools and AST data types are used to handle parsing, type-checking, type environment inspection, and pretty printing of modified code. Modified code is further beautified by running it
through \verb+ocamlformat+~\cite{OCamlFormat} if the user has it installed. Comments are (unfortunately) discarded by OCaml's parser.

\begin{figure}[t]
  \centering
  \includegraphics[width=\textwidth]{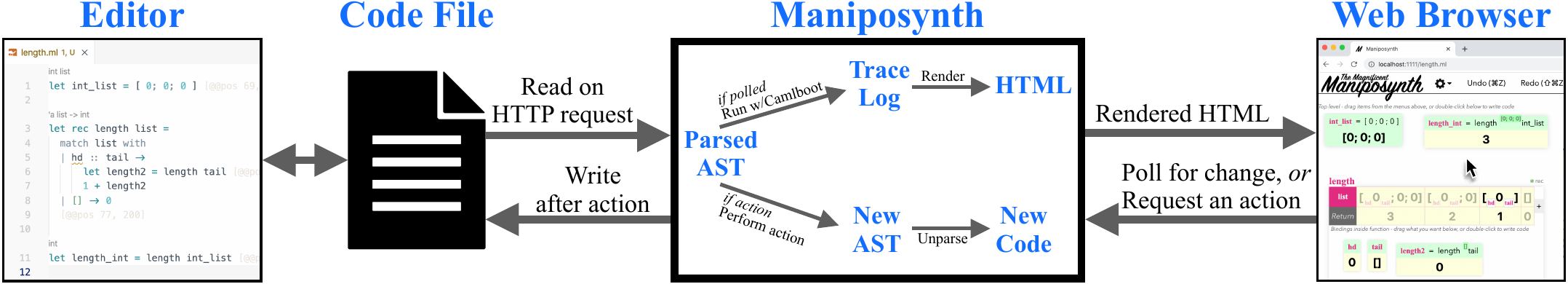}
  \caption{\ms{} architecture overview.}
  \label{fig:architectureOverview}
\end{figure}

For displaying live feedback, we need to run the program and log the runtime values flowing through the code. We modified the OCaml interpreter from the Camlboot~\cite{camlboot} project to emit a trace of all runtime values at all execution steps.  We also performed additional modifications to handle holes and assertions (described in the next section).

After \ms{} runs the code via our modified Camlboot, \ms{} associates runtime values from the logged execution trace with program expressions, and then renders HTML which is displayed in the browser. Almost all OCaml-specific logic is handled server-side. The client-side Javascript only handles TV positioning and standard GUI interaction logic. When the user performs an action, the Javascript sends the action to the server via HTTP, the code is modified on disk, and the server prompts the browser to reload the page to re-render the display. The browser also polls the server via HTTP so that when the file is changed on disk, the display will refresh. This overall architecture is outlined in \autoref{fig:architectureOverview}.

Below, we describe our modified Camlboot interpreter, then how bindings are reordered to provide a non-linear experience, and lastly the mechanics of the synthesizer.

\subsection{Interpreter}

\ms{} needs to provide live runtime values.
We base \ms{} on the OCaml interpreter in the Camlboot~\cite{camlboot} project, an experiment in bootstrapping the OCaml compiler. The Camlboot OCaml interpreter is written in OCaml and represents all values as an ordinary OCaml algebraic data type (ADT), which allows inspecting their type and structure at runtime, at the cost of somewhat slower execution relative to compiled code. We modified Camlboot to handle holes and assertions, and to log runtime values during execution.

\newcommand{\synthForm}[1]{{\color{NavyBlue} #1}}
\newcommand{\synthFormColorShort}{blue}

\begin{figure}[b]
\newcommand{\termOr}{\ |\ }
\newcommand{\kw}[1]{\text{\textbf{#1}}}
\newcommand{\additionalGrammerLine}[1]{\\ & $\termOr$ & $#1$}
\newcommand{\additionalGrammerLineWithLabel}[2]{\\ \textbf{#1}\hspace{0.03in} $\phantom{B}$ & $\termOr$ & $#2$}
\newcommand{\grammerLine}[3]{\textbf{#1}\hspace{0.03in} $#2$ & $::=$ & $#3$}
\newcommand{\grammerKindSep}{\vspace{0.03in}\\}
\centering
{
\renewcommand{\arraystretch}{0.9} 
\begin{minipage}{.5\textwidth}%
\begin{tabular}{rll}
\grammerLine{Programs}{P}{\overline{\kw{type}\ t\ =\ T}\ \ \overline{B}}

\grammerKindSep{}

\grammerLine{Types}{T}{\text{(from OCaml)}}

\grammerKindSep{}

\grammerLine{Top-Level}{B}{\kw{let}\ x = e}
\additionalGrammerLineWithLabel{Bindings}{\kw{let rec}\ x_1 = e_1} 
\additionalGrammerLine{\kw{let}\ () = \kw{assert}\ (e_1 = e_2)}

\grammerKindSep{}

\grammerLine{Patterns}{p}{\synthForm{C} \termOr \synthForm{C\ x} \termOr \synthForm{C\ (x_1\overline{,x_i})}}

\end{tabular}

\end{minipage}%
\begin{minipage}{.5\textwidth}%

\begin{tabular}{rll}
\grammerLine{Exp.}{e}{(??) \termOr \synthForm{c} \termOr \synthForm{C} \termOr \synthForm{C\ e} \termOr \synthForm{C\ (e_1\overline{,e_i})}}
\additionalGrammerLine{\synthForm{x} \termOr \synthForm{\kw{fun}\ x \rightarrow e} \termOr e_1\ \overline{e_i} \termOr \synthForm{x\ \overline{e_i}}}
\additionalGrammerLine{\kw{let}\ x = e_1\ \kw{in}\ e_2} 
\additionalGrammerLine{\kw{let rec}\ x_1 = e_1\ \kw{in}\ e_b} 
\additionalGrammerLine{(e_1, e_2\overline{, e_i})}
\additionalGrammerLine{\synthForm{\kw{if}\ e_1\ \kw{then}\ e_2\ \kw{else}\ e_3}}
\additionalGrammerLine{\synthForm{\kw{match}\ e_1\ \kw{with}\ \overline{p \rightarrow\phantom{t}e_i}}}
\end{tabular}
\end{minipage}
}

\caption[The OCaml subset fully supported by \ms{}]{The subset of OCaml fully supported by \ms{}. Overlines denote zero or more instances of the syntactic element. Unsupported expressions and patterns are displayed but do not have full UI support. The synthesizer (\autoref{sec:synthesizer}) emits only those forms displayed in \synthForm{\synthFormColorShort{}}.}
\label{fig:supportedSubset}
\end{figure}

\subparagraph*{Supported Subset}

Unmodified, the Camlboot interpreter will run a large subset of OCaml. The tooling and display in \ms{}, however, currently only fully supports a smaller subset, shown in \autoref{fig:supportedSubset}. At the top-level, programs in \ms{} are expected to consist only of type declarations followed by (potentially recursive) let-bindings; assertions are expected to occur only at the top-level. Only single-name patterns have full UI support (although internal operations such as free variable analysis account for nested patterns). Supported expressions include holes, base value constants, argument-less, single-argument, and multi-argument constructors, variable usages, function introductions with an unlabeled parameter, multi-argument function applications, (potentially recursive) let-bindings, tuples, if-then-else, and pattern match case splits. Case splits are only fully supported on constructors.

Records do not have complete UI support. User-defined modules, opening modules, imperative functions, and object-oriented features are currently unsupported.

The swath of supported syntax was enough to cover the kinds of data structure manipulations we explored in our evaluation. During the user study exercises, participants rarely missed the unsupported syntax. Even so, for the tool to become practical for everyday use, the users noted it would need to support modules and imperative programming.

\subparagraph*{Holes and Bombs} It is best for the user if live feedback is available even if the program is incomplete. While we could have the interpreter crash on the first hole, that may still be too restrictive, \eg{} if the expression is new and is still dead code, then the presence of the hole should be inconsequential to the rest of execution. A thorough solution would be to adopt the Hazelnut Live semantics, which describes how to evaluate \emph{around} holes~\cite{HazelnutLive}. When holes reach elimination position, terms become stuck (\eg{} What should hole plus hole be? Or which case branch should we take when the scrutinee is a hole?). Hazelnut Live evaluates around the term by, effectively, turning the stuck term into a value which is propagated until it causes another term to become stuck, and so on. While this can offer intriguing UI possibilities in its own right (outlined in~\cite{HazelnutLive}), it requires displaying stuck terms to users as if they are values. \ms{} may do so eventually, but our display is already full of elements to keep track of. Asking users to make sense of stuck terms, displayed far from their origin, might be confusing.

\ms{} adopts a middle ground that does not allow terms to become stuck. The hole expression \verb+(??)+ introduces a hole value \verb+?+ that remembers the introduction location and captures a closure. (This closure is not displayed, but is occasionally used during synthesis.) Hole values propagate through evaluation. If a hole value reaches elimination position (\eg{} \verb-? + ?-), we resolve the expression to a special Bomb value (displayed as \includegraphics[height=0.8em]{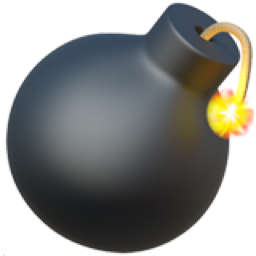}). Similarly, if a Bomb reaches elimination position, another Bomb is produced. In this way, execution continues and expressions unrelated to the unfinished code still provide live feedback.

Finally, to prevent infinite loops from inhibiting live feedback, \ms{} uses fueled execution to abort when the right-hand side of a binding takes too long to execute. Each top-level let-binding is allocated 1000 units of fuel (execution steps), and each non-top-level let-binding reserves 50 units for later execution in case the binding diverges. When the interpreter runs out of fuel on a binding, all patterns at the binding are bound to Bomb, and execution continues if any fuel remains. Divergence is moderately common, because recursive call skeletons like \verb+length (??)+ from the \nameref{sec:overview} will repeatedly call the function with a hole value. Thus it is important that execution bypasses the divergence with some fuel reserved so that later TVs will still show live values in the display.

\subparagraph*{Assertion Logging} When an assertion is encountered during execution, ordinarily OCaml would throw an exception if the asserted expression returns false. Instead, \ms{} logs the result for later, but never raises an exception. Only equality comparisons are supported for now; unsupported assertions are skipped. The expected expression (the right-hand side), the subject expression (the left-hand side), and the result values of both are logged. Logged assertions are used both for synthesis and to display blue expected values to the user wherever that same expression and value is encountered during execution (\autoref{fig:assertsAndIncorrectSynthResult}a).

\subparagraph*{Tracing} In addition to assertions, \ms{} also logs other execution information needed for display. Our interpreter records information in two places: each execution step is entered into a global log, and we tag side information onto runtime values.

At each execution step and at each pattern bind we add a log entry to a global trace, recording the current AST location, the call frame number (from a global counter incremented upon each function call), the result value or value being pattern matched against, and the execution environment of bound variables. When producing the display, this information is queried to discover which values flowed through which locations and under which call frames, and the appropriate values are rendered.

For convenience, we also store extra information on values. On values we log the type of the value when it is introduced (or returned from a built-in external primitive such as addition) so we have a concrete type associated with the value even if it is later used in a polymorphic context. To the value we also attach a list of frame numbers and AST locations of the expressions and patterns the value passes through, to, \eg{}, conveniently interrogate where a value was first introduced. For example, to display function closure values, we find where the closure was bound to a name and display that name as the rendered value.

These mechanisms are sufficient to render the live feedback in \ms{}. Extensive logging might be expected to slow down execution. But when applied to the small programs tested at present, HTML rendering tends to take longer than the initial execution. Overall, the \ms{} server is generally able to provide a response in under 200ms.

\subsection{Fluid Binding Order}
\label{sec:fluidBindingOrder}

A primary goal of \ms{} is to offer a non-linear editing experience. We do not want users to have to worry about binding ordering. If the user sees a name on the 2D canvas, they should be able to reference that name in the expression they are editing even if, in the written code, that name is introduced later in the program.

\begin{figure}[t]
  \begin{tabular}{ccc}
  \includegraphics[width=0.305\textwidth]{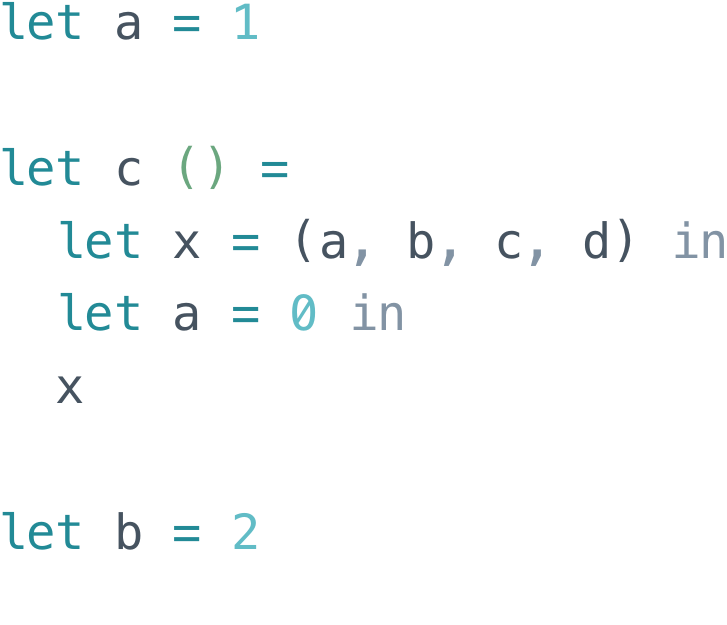} &
  \includegraphics[width=0.305\textwidth]{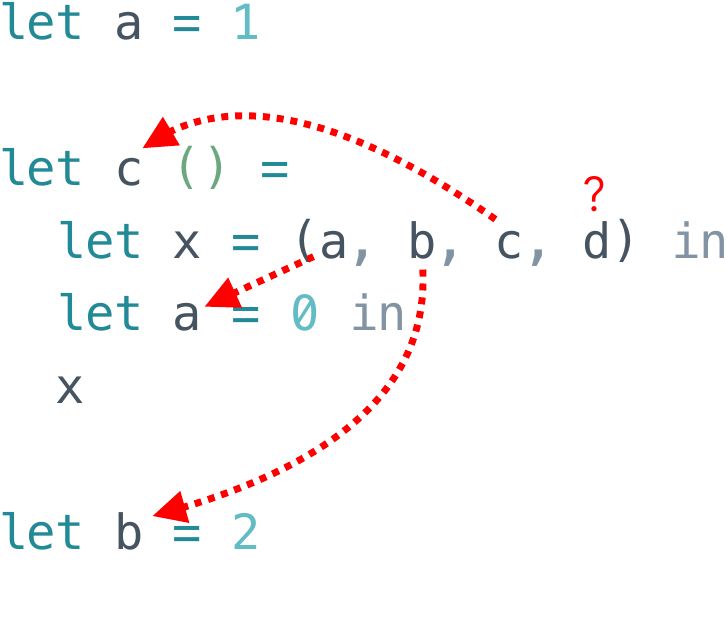} &
  \includegraphics[width=0.305\textwidth]{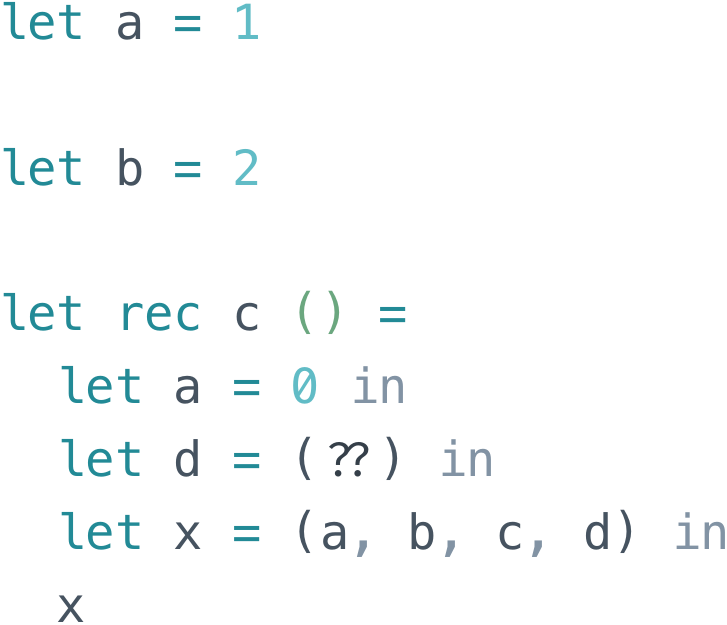}
  \\
  (a) & (b) & (c)
\end{tabular}

  \caption{(a) A program. (b) How \ms{} interprets what each variable usage refers to. (c) The reordering \ms{} performs. The bindings of \texttt{a} and \texttt{b} are moved up so they are in scope, \texttt{c} is marked as recursive, and \texttt{d} is created.}
  \label{fig:reorderingExample}
\end{figure}

To support this non-linear workflow, only limited variable shadowing is supported. All names are assumed to be unique within each top-level definition. Names may (initially) be out of order. \autoref{fig:reorderingExample}a presents an example. As shown in \autoref{fig:reorderingExample}b, \ms{} interprets each of the variable uses in \verb+(a, b, c, d)+ different from standard OCaml. Although there is a top-level binding \verb+a = 1+ in scope, the use of \verb+a+ in \verb+(a, b, c, d)+ refers to the binding \verb+a = 0+ below it, because \ms{} is agnostic about the order of names within a top-level definition (within the top-level definition \verb+c+). The variable \verb+b+, however, does not have a definition within \verb+c+, so its usage refers to the top-level binding \verb+b = 2+, despite being (momentarily) out of scope. Similarly, the usage of \verb+c+ refers to the top-level binding for \verb+c+, although the binding is not (yet) marked recursive. The variable \verb+d+ has no binding.

To reify these references as valid OCaml, after every user action \ms{} reorders bindings, move bindings into \verb+match+ statement branches (not shown here), and adds a \verb+rec+ flag on bindings that refer to themselves. Only single recursion can be inferred for now (multiple recursion may be added manually in the text editor). \autoref{fig:reorderingExample}c shows the result for this code example. The nested definition of \verb+a+ is moved before \verb+x+, the top-level definition of \verb+b+ is moved up as well, the binding for \verb+c+ is marked as recursive, and a new definition is added for the missing variable \verb+d+.

When creating a new definition for an undefined variable, if it is used as a function in an application the new variable is bound to a function skeleton with the appropriate number of parameters.
Here, \verb+d+ is not used as a function and is initialized as a simple hole \verb+(??)+.

The overall effect of the above reordering is that users rarely need to think about binding order in their code. They can use the displayed TVs as if they are all visible to each other.

\subsection{Synthesizer}
\label{sec:synthesizer}

As discussed in the \nameref{sec:overview}, \ms{} includes a programming by example (PBE) workflow to help users finish their incomplete code. Here we detail the program synthesizer's operation and our design choices in its implementation.

The \ms{} synthesizer does not contain any new ``big'' ideas, but the design was carefully chosen for our setting. To be as practical as possible, we had four goals:

\begin{alphaenumerate}
  \item \textbf{Few examples.} To reduce user burden, we would like the synthesizer to operate with few examples. For example, \textsc{Myth}~\cite{Myth} also targeted an OCaml subset, but required that user-provided examples include all needed recursive calls—\eg{} \verb+length [0;0] = 2+, \verb+length [0] = 1+, and \verb+length [] = 0+. This ``trace completeness'' requirement is burdensome; we would like our synthesizer to operate with only one or two examples.
  \item \textbf{No type annotations.} Similarly, \textsc{Myth} and its successor \textsc{Smyth}~\cite{Smyth} require holes to have types before synthesis, which requires manual annotation. We would like to relieve the user of this responsibility and operate without explicit type annotations.
  \item \textbf{As simple as possible.} The primary goal of \ms{} is to explore non-linear editing, not synthesis per se, so we wanted to keep our synthesizer as simple as possible. For now, we did not adapt \textsc{Smyth} because, although it appropriately relaxes the trace-completeness requirement, \textsc{Smyth} utilizes a complicated synthesis schedule and requires the Hazelnut Live machinery~\cite{HazelnutLive} for evaluating around holes. 

  \item \textbf{Quality results.} When given only a few examples, synthesizers are notorious for producing simple but undesirable results (for example, ``January, Febuary, Maruary''~\cite{FlashFillWrongPattern}) which limits their utility. This problem is compounded when the synthesizer is asked to operate in practical environments with many variables in scope, rather than unrealistic bare minimal execution environments often used for synthesizer benchmarks. Our synthesizer should operate with the OCaml standard \verb+Pervasives+ library open in the execution environment so the synthesizer may choose to use, for example, addition and subtraction. We adopt statistics and heuristics to make this tractable.
\end{alphaenumerate}

\textsc{Myth} used types and examples to dramatically reduce the search space and to intelligently introduce case splits. To meet the above goals, we built a \textsc{Myth}-like synthesizer, but we relax the trace-completeness requirement and instead rely on a statistics model to guess more likely terms sooner. Our target language subset, the statistics model, the synthesis operation, and our other heuristic choices are described below.

\subparagraph*{Synthesizable Subset} The \synthForm{\synthFormColorShort} expression and pattern forms in \autoref{fig:supportedSubset} describe the OCaml subset that the synthesizer may produce to fill holes in the program. It can introduce functions, \verb+match+ expressions, constants (drawn from a corpus), variable uses, function calls with a variable in function position, constructor uses, and if-then-else expressions.

\subparagraph*{Statistics Model} Naively, guess-and-check will produce a large number of programs the user is unlikely to want. Incorporating a statistics model guides the synthesizer to guess more likely programs sooner and can speed up synthesis by multiple orders of magnitude~\cite{Euphony,MaxFlash}. It also has the potential to offer more reasonable results when fewer examples are given.

We model program likelihood using a probabilistic context-free grammar (PCFG). A PCFG is a grammar with a probability assigned to each production rule. For our synthesizer, we derived the probabilities of the production rules from a corpus of OCaml code—namely, the source files required to build the OCaml native compiler. The overall probability of a program term is the product of the probability of the production rule of the term with (recursively) the probability of all its subterms. Identifiers are handled specially: an identifier's probability is based on how spatially close it is to its introduction pattern in the code.

For example, the most likely term (\ie{} what the synthesizer should guess first) is always the most recently introduced variable. The production rule for an identifier has a probability of 52\%, the probability that the identifier is local is 73\%, and the probability a local identifier is the most recently introduced variable is 31\%, for an overall probability of 12\%.

\subparagraph*{Type-Directed Refinement} \textsc{Myth} divides synthesis into two processes. \emph{Type-directed refinement} introduces program sketches—either data constructor applications, function introductions, or case splits—at holes based on the type at the hole and the types of variables in scope (to find an appropriate scrutinee for a \verb+match+). These sketches contain further holes to fill (\ie{} for the function body and the match branches). Type-directed refinement alternates with \emph{type-directed guessing}, which performs simple type-constrained term enumeration to fill remaining holes (guessing will not introduce functions or \verb+match+ statements).

\ms{} uses type-based refinement to introduce functions and insert case splits (data constructors are instead handled in the guessing process). \ms{} refines a hole into a function zero to three times (\ie{} supporting up to three arguments) before possibly introducing a single case split. The user can add further case splits with the ``Destruct'' button. Introducing functions is rarely needed in practice because, in the \ms{} UI, undefined variables are inserted with a function skeleton.

\subparagraph*{Type-Directed Guessing}
After refinement, terms are enumerated (guessed) at holes up to a probability bound~\cite{BranchAndBound} according to the PCFG. During term enumeration, the probability bound is treated as a resource. When the probability is exhausted, no further enumeration occurs on a subtree.
If a candidate subterm's probability is above the final bound, the remaining probability is available for enumerating sibling terms.

Within a hole, term enumeration is type-directed, starting from the type of the hole. Leveraging OCaml's type checking machinery, subterms are unified during the enumeration process to narrow the type. For example, if a hole has type \verb+int+ and the synthesizer guesses a call to \verb+max+, of type \verb+'a → 'a → 'a+, the return type will be unified with \verb+int+ and the synthesizer will only guess terms of type \verb+int+ for the arguments.

Initial sketches often have polymorphic types unhelpful for synthesis. For example, the \verb+length+ function has type \verb+'a → 'b+ in \verb+let length x1 = (??)+.
To tighten these bounds before term enumeration, input and output types of functions are speculatively chosen based on the examples. If the user asserts that \verb+length [0] = 1+, then \verb+length+ is given the speculative type \verb+int list → int+.
Speculative types are removed after term enumeration in case the final code has a more general inferred type (\verb+'a list → int+ in this case).

To produce more natural results, \ms{} limits where constants may appear. A term is estimated to be non-constant if it uses an introduced function parameter or variable introduced under the outermost enclosing function. At most one hole may be constant, and, when introducing a function call, at least one argument must be non-constant. Term enumeration avoids constants in locations where inserting one would violate these rules.

\subparagraph*{Non-Linearity} To follow the non-linear behavior of \ms{}, the guessing process also guesses variable names that are not in lexical scope but could be moved into scope via \ms{}'s binding reordering algorithm (\autoref{sec:fluidBindingOrder}). The reordering algorithm is applied before testing whether a candidate program satisfies all assertions.

\subparagraph*{Final Heuristics}
Finally, when all holes have been filled with type-appropriate terms within the probability bound, the candidate program is accepted if:

\begin{alphaenumerate}
  \item All assertions are satisfied. (Fueled execution is used when checking assertions.)
  \item At most one hole is filled with a constant.
  \item All introduced function parameters are used.
  \item The result at a hole has not previously been rejected by the user.
  \item Execution of the examples encounters all filled hole locations (\ie{} the execution path does not somehow avoid a hole).
\end{alphaenumerate}

If no satisfying hole fillings are found at the initial probability bound and a 10 second timeout has not been reached, guessing is restarted with a new bound 1/20 of the old. If there is a valid candidate program, the program with highest probability is returned. Enumeration within a given probability bound is not precisely from highest to lowest probability, however, so a timeout will not interrupt a round of synthesis until the full space of that probability bound is explored. Thus, the timeout the user experiences varies between 10 and 40 seconds.

\section{Evaluation}
\label{sec:evaluation}

To evaluate to what degree \ms{} meets its goal of providing value-centric, non-linear editing, we performed two evaluations. In one, an expert user (the first author) used \ms{} to implement 38 functions from the exercises and homework of a functional data structures course~\cite{IN2347}. In the second, to provide additional qualitative insights on the operation of the tool, we hired two professional OCaml programmers, guiding and observing them as they used \ms{} to implement a subset of the above functions.

\subsection{Study Setups}

The first six lessons of the course~\cite{IN2347} cover natural numbers (via an ADT), various list functions, leaf trees, binary trees, binary search trees, and a form of binary search tree that also records on each node the minimum value of all its descendants. We excluded the six functions on this specialized tree because of time constraints. The course exercises and homework spanned 38 functions on the remaining data structures. The first author implemented each of these functions in \ms{} with the code editor hidden. \ms{} was configured to log the number and kinds of actions performed. We report on these in the next section. Our aim was to show that the \ms{} interface was able to implement these exercises and to discover if there were any obvious trouble points.

For our user study, we advertised on \url{https://discuss.ocaml.org/} and hired two professional OCaml programmers for three sessions each. Sessions were spread over three weeks, with each session lasting two hours. Participant 1 (P1) and Participant 2 (P2) had 5 and 11 years, respectively, of professional OCaml experience. The participants ran \ms{} on their own computers alongside their preferred text editor (Vim for both). The study facilitator connected via video conference and recorded the sessions. Participants implemented their choice of exercises from the list, or suggested their own task to complete. The facilitator provided varying amounts of guidance throughout, starting with close guidance to teach the tool and transitioning to less intervention as participants became more comfortable. After each exercise and at the end of each session, participants discussed a series of questions posed by the facilitator. In concert with \ms{}'s four design principles—value-centric operation, non-linearity, supporting synthesis, and bimodality—we aimed to gain insights about the following four research questions, along with three supplemental questions:

\begin{description}
  \item[RQ1.] How do users interact with the live values?
  \item[RQ2.] How do users work non-linearly?
  \item[RQ3.] How do users interact with program synthesis?
  \item[RQ4.] How do users interact with their text editor?
  \item[SQ1.] What are the pain points? How might the system be improved?
  \item[SQ2.] Are participants comfortable enough to complete a task without guidance?
  \item[SQ3.] What else can we learn via the lenses of the Cognitive Dimensions of Notations~\cite{CognitiveDimensions}?
\end{description}

For each participant, the first session introduced the tool without synthesis, the second session introduced synthesis (before the synthesizer had a statistics model), and the third session concluded with the tool as presented here. Based on participant feedback, we fixed bugs and made improvements between each session.

\begin{figure}[bt]
  \centering
  \includegraphics[height=1in]{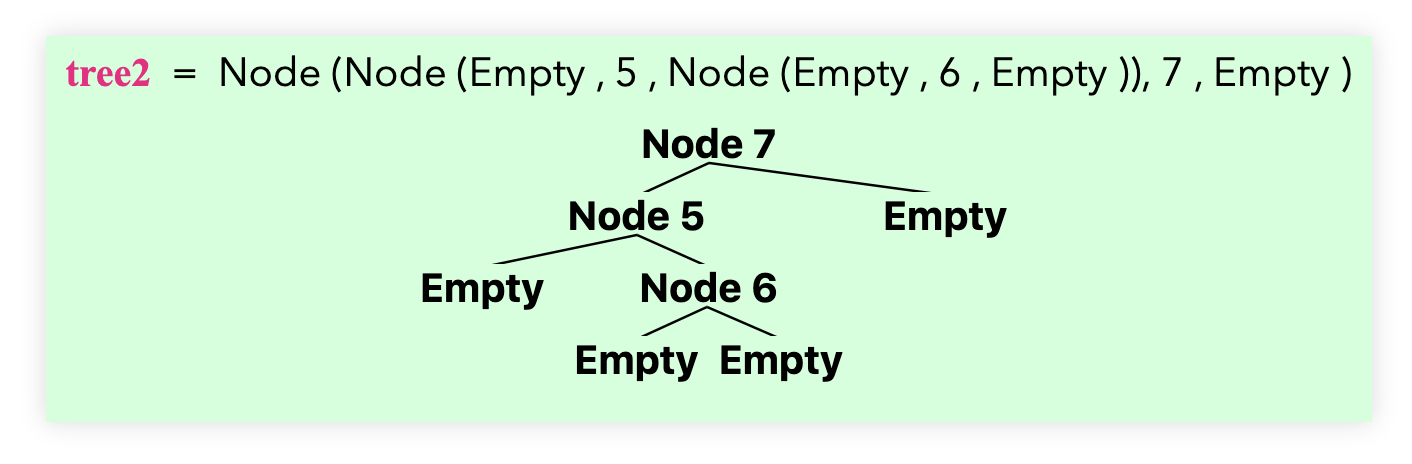}
  \caption{\ms{} beautifies tree-like values.}
  \label{fig:tree}
\end{figure}

\subsection{Results}

\newcommand{\tableFontSize}[1]{{\scriptsize #1}}
\newcommand{\tableRow}[9]{
  \tableFontSize{#1} &
  \tableFontSize{#2} &
  \tableFontSize{#3} &
  \tableFontSize{#4} &
  \tableFontSize{#5} &
  \tableFontSize{#6} &
  \tableFontSize{#7} &
  \tableFontSize{#8} &
  \tableFontSize{#9} \\
}

\begin{table}[hbt]
\scriptsize
\centering
\setlength\tabcolsep{3pt}
  \begin{tabular}{l|cccccccc}
\tableRow
  {\textbf{Function}}
  {\textbf{LOC}}
  {\textbf{Asserts}}
  {\textbf{Time}}
  {\textbf{Mouse}}
  {\textbf{Keybd}}
  {\textbf{Un/Re/Del}}
  {\textbf{TypeErr}}
  {\textbf{Crash}}
\hline
\tableRow{nat\_plus          }{ 5  }{   }{ 0.8  }{ 6  }{ 5   }{    }{   }{   }
\tableRow{nat\_minus         }{ 8  }{   }{ 1.9  }{ 6  }{ 11  }{    }{   }{   }
\tableRow{nat\_mult          }{ 9  }{   }{ 1.4  }{ 8  }{ 6   }{    }{   }{   }
\tableRow{nat\_exp           }{ 13 }{   }{ 2.1  }{ 9  }{ 6   }{    }{   }{   }
\tableRow{nat\_factorial     }{ 13 }{   }{ 1.6  }{ 8  }{ 4   }{    }{   }{   }
\tableRow{nat\_map\_sumi     }{ 10 }{   }{ 2.6  }{ 11 }{ 5   }{    }{ 1 }{   }
\tableRow{count              }{ 9  }{   }{ 1.9  }{ 9  }{ 11  }{    }{   }{   }
\tableRow{length             }{ 4  }{   }{ 0.3  }{ 1  }{ 7   }{    }{   }{   }
\tableRow{snoc               }{ 8  }{ 1 }{ 2.4  }{ 8  }{ 12  }{ 2  }{   }{   }
\tableRow{reverse            }{ 8  }{   }{ 1.5  }{ 4  }{ 9   }{    }{   }{   }
\tableRow{nat\_list\_max     }{ 17 }{   }{ 4.6  }{ 23 }{ 21  }{    }{   }{   }
\tableRow{nat\_list\_sum     }{ 13 }{   }{ 1.1  }{ 9  }{ 4   }{    }{   }{   }
\tableRow{fold               }{ 9  }{   }{ 3.2  }{ 14 }{ 6   }{    }{   }{   }
\tableRow{shuffles           }{ 14 }{   }{ 14.5 }{ 25 }{ 28  }{ 2  }{   }{   }
\tableRow{contains           }{ 9  }{   }{ 2.2  }{ 10 }{ 13  }{ 1  }{   }{   }
\tableRow{distinct           }{ 16 }{   }{ 2.4  }{ 9  }{ 11  }{ 2  }{   }{   }
\tableRow{foldl              }{ 10 }{ 1 }{ 1.5  }{ 10 }{ 6   }{    }{ 1 }{   }
\tableRow{foldr              }{ 8  }{ 1 }{ 1.8  }{ 10 }{ 5   }{    }{   }{   }
\tableRow{slice              }{ 12 }{ 3 }{ 9.8  }{ 19 }{ 22  }{ 4  }{   }{   }
\tableRow{append             }{ 8  }{ 1 }{ 1.4  }{ 7  }{ 9   }{    }{   }{   }
\tableRow{sort\_by           }{ 21 }{ 3 }{ 6.2  }{ 17 }{ 29  }{    }{   }{   }
\tableRow{quickselect        }{ 13 }{ 1 }{ 13.1 }{ 19 }{ 38  }{ 1  }{ 1 }{   }
\tableRow{sort               }{ 16 }{ 3 }{ 5.6  }{ 11 }{ 32  }{ 2  }{   }{   }
\tableRow{ltree\_inorder     }{ 12 }{ 1 }{ 2.9  }{ 7  }{ 20  }{ 1  }{ 1 }{   }
\tableRow{ltree\_fold        }{ 13 }{ 1 }{ 3.1  }{ 13 }{ 13  }{    }{   }{   }
\tableRow{ltree\_mirror      }{ 11 }{ 1 }{ 4.4  }{ 12 }{ 6   }{    }{ 1 }{ 1 }
\tableRow{bst\_contains      }{ 14 }{ 3 }{ 6.6  }{ 11 }{ 32  }{ 1  }{   }{   }
\tableRow{bst\_contains2     }{ 17 }{ 5 }{ 10.4 }{ 20 }{ 41  }{ 2  }{   }{   }
\tableRow{btree\_join        }{ 34 }{ 2 }{ 61.7 }{ 82 }{ 64  }{ 51 }{   }{ 2 }
\tableRow{bst\_delete        }{ 36 }{ 2 }{ 14.4 }{ 31 }{ 24  }{ 4  }{   }{   }
\tableRow{bstd\_valid        }{ 29 }{ 3 }{ 32.2 }{ 63 }{ 100 }{ 4  }{ 1 }{   }
\tableRow{bstd\_insert       }{ 18 }{ 2 }{ 8.0  }{ 38 }{ 23  }{ 3  }{   }{   }
\tableRow{bstd\_count        }{ 21 }{ 1 }{ 7.6  }{ 15 }{ 32  }{ 1  }{   }{   }
\tableRow{bst\_in\_range     }{ 31 }{ 3 }{ 9.3  }{ 23 }{ 39  }{ 3  }{   }{   }
\tableRow{btree\_enum        }{ 29 }{ 3 }{ 19.2 }{ 31 }{ 51  }{ 6  }{ 3 }{   }
\tableRow{btree\_height      }{ 15 }{ 1 }{ 1.9  }{ 11 }{ 14  }{    }{   }{   }
\tableRow{btree\_pretty      }{ 14 }{ 1 }{ 3.7  }{ 4  }{ 21  }{    }{ 4 }{   }
\tableRow{btree\_same\_shape }{ 19 }{ 1 }{ 8.1  }{ 14 }{ 34  }{ 7  }{   }{   }
\hline
\tableRow
  {\textbf{Total}}
  {\textbf{566}}
  {\textbf{44}}
  {\textbf{277.6}}
  {\textbf{628}}
  {\textbf{814}}
  {\textbf{97}}
  {\textbf{13}}
  {\textbf{3}}
	\end{tabular}

  \vspace{0.05in}

  \caption[Examples]{Example exercises, with lines of code, number of assertions, time in minutes, number of mouse actions (excluding selection and undo/redo), number of keyboard interactions (\eg{} typing in a textbox), number of undo/redo/deletions, number of type errors encountered, and number of times \ms{} crashed and the file had to be repaired in the text editor.}
  \label{table:examples}
\end{table}

\subparagraph*{Example Corpus Implementation} The expert implementer spent about 4.5 hours total---working in fullscreen with their text editor hidden---implementing the 38 functions, resulting in about 550 lines of code (including AST annotations and examples written for live feedback, but excluding whitespace). A quantitative summary of these example exercises is shown in \autoref{table:examples}. There are often many paths to a correct implementation, so to constrain the workflow the implementer did not use synthesis, and did not use the ordinary text editor except to copy an earlier function into a later exercise (in case of dependencies) or when \ms{} crashed on the given code (\eg{} when the implementer tried to raise an exception in unreachable code that turned out to be reachable!). Note that for the functions operating on tree-like data types (ADTs with multiple children of the same recursive type), \ms{}'s live display helpfully draws the trees as trees (\autoref{fig:tree}).

Primarily, these 38 examples show that the \ms{} UI is expressive enough to create these programs. We also noticed two qualitative takeaways from the exercises. First, although we believe bimodality is an important property for the grounding and long-term practicality of the tool, it is possible to hide the text editor and work entirely in \ms{}. Second, even with live feedback available, it is not always used—a theme that reemerged in our user study. We now discuss these two observations in more detail.

The non-linearity machinery largely worked—the implementer did not have trouble with binding order. Even so, they were careful to name extracted subvalues well, because the positioning of the extracted TVs on the 2D display did not (by default) reflect the items' positions in the original data structure. Particularly for nested matches, there were sometimes a large number of these extracted values displayed and it was hard to keep track of them. The implementer found it helpful to reposition the extracted TVs (the TVs representing case split branch patterns) to reflect those original positions. Ordinary textual code for case split patterns would provide some of these positional cues without manual interaction.

On a few exercises requiring nested \verb+match+ statements, \ms{} initially created the wrong nested \verb+match+ structure; with the non-linear display, this is hard to notice and requires thinking about the \verb+match+ nesting structure shown in the return TVs area. Regardless, the implementer worked around the trouble by undoing and triggering the destructions differently.

The second observation from these examples is that the implementer noticed that they seem to flip between two mental modes: these modes correspond roughly to focusing on displayed values versus focusing on expressions. In the \emph{value-oriented} mode, the implementer would put their attention on the live values to consider if the code is operating correctly; in the \emph{expression-oriented} mode, the implementer would read expressions and simulate the computer's operation in their head.
As a matter of discipline, the implementer was trying to push themselves to consider and use the live values, but nevertheless often reverted to thinking only about the expressions instead. We have three hypotheses for why there seems to be a tendency to revert to focusing on expressions instead of values.

\emph{Hypothesis A:} Expressions are a concise language to represent abstractions, and programming is, fundamentally, abstract. Concrete values do not directly represent the abstraction.

\emph{Hypothesis B:} Seasoned programmers have many years of experience reading code and simulating the computer in their head. Our brains have adapted to it, and it feels natural.

\emph{Hypothesis C:} \ms{} did not provide enough live feedback and forced the implementer to consider the expressions. In some cases this was immediately true: \ms{} currently only displays the first and last three call frames, with no option to see the others, so sometimes relevant values were in unavailable call frames. Additionally, when initially trying to figure out what algorithm was needed at all, the implementer found it easier to work out the initial algorithm sketch in their head rather than guess and check in \ms{}.

All three reasons likely contributed to the tendency to put attention back on expressions rather than values. A similar theme was observed in the user study, which we now discuss.

\subparagraph*{RQ1. How do users interact with the live values?}
Value-oriented focus versus the ``old way'' of expression-oriented focus is a theme that appeared in several participant interactions. For example, despite values featuring prominently in the display, it took until after the entire first exercise for P2 to fully realize they were looking at and working with live \emph{values}. In another scenario, P1 and the facilitator together spent an embarrassingly long time trying to find a bug in an \verb+insert_into_sorted_list+ helper. After finding the bug they realized that, had they carefully inspected the live values, they might have found the bug much sooner. Additionally, P2 observed that they are so used to reading trees as long lines of text (\eg{} \verb+Node (Leaf 2, Node (Leaf 2, Leaf 3))+) that they were subtly repelled by the ``much more readable'' beautified 2D rendering of tree values (\autoref{fig:tree}).

Even so, participants still did use the live display and expressed appreciation for it. For example, P2 noted that the live display ameliorated the need to write large amounts of tree pretty-printing code to perform printf-debugging.

Live values require the user to switch call frames to see other example function calls, or calls that hit a different branch in the code. This was not always natural for participants. In the first session, P2 admitted to sometimes being confused about what branch they were looking at. And, despite gaining moderate proficiency with the tool by the end of the study, P2 still remarked that it was hard to think about how you can flip between frames. How to modify the display to help clarify this operation remains an open question.

\subparagraph*{RQ2. How do users work non-linearly?}
We want to know how programmers adapt to \ms{}'s non-linear style. The tool requires a number of ``inside-out''  (P1) changes in thought, such as creating an example before defining a function, providing expressions \emph{without} naming them first, and not worrying about let-binding order.
By the end of the study, participants were familiar with these concepts but did not necessarily start out that way. For example, in the first session P1 had trouble remembering to create functions by first providing an example call, but by the end of the study was doing so without any prompting from the facilitator. P1 also initially had trouble finding particular variable definitions on the screen but felt more comfortable by the second session. Near the end of the second session P2 expressed, ``I want a let binding…I don’t have any confidence I can make let bindings,'' despite having successfully done so many times by double-clicking the subcanvas or dragging values into the subcanvas. P2 instantly understood after a quick reminder from the facilitator, but it is notable that even after around three hours with the tool it had not quite sunk in that most TVs are let-bindings.

At the end of the study, we asked participants about writing expressions without naming them first. P1 expected to prefer being required to always provide a name; P2 was unsure, but noted that \ms{}'s default names had improved from the first version they used. In particular, at P2's behest we hard-coded the default names for list destruction to be \verb+hd::tail+ instead of the original type-based \verb+a2::a_list2+. Even so, we rediscovered that naming was important in programming. Function skeletons are still inserted with generic parameter names, \eg{} \verb+fun x2 x1 -> (??)+, which are both unhelpful and backwards. This indeed resulted in user mistakes, and is a point to improve in future versions of \ms{}.

Despite a few troubles, both participants were positive overall about the non-linear workflow. P1 noted the non-linear style ``fits a lot more with how I like to write code,'' and P2 said, ``I like it, I’m excited about it.''

\subparagraph*{RQ3. How do users interact with program synthesis?}
We introduced participants to the synthesizer in the second session, at which point it lacked a statistics model (instead enumerating terms by size) and did not offer ``Accept / Reject'' buttons (instead requiring the user to Undo upon undesired results); these were added for the final session. We wanted to learn how comfortable users were with providing assertions and using the synthesizer.

\begin{table}[bt]
\centering
  \begin{tabular}{l|ccccccc}
& & & \textbf{Mean} & \textbf{Mean} & \textbf{\% No} & \textbf{\% Useless} & \textbf{\% Useful} \vspace{-3pt}\\
\textbf{Participant} & \textbf{\#Ex} &\textbf{\#Synth} & \textbf{\#Assert} & \textbf{\#Hole} & \textbf{Result} & \textbf{Result} & \textbf{Result} \\  \hline
P1 & 7 & 52 & 3.3 & 2.0 & 48\% & 35\% & 17\% \\
P2 & 7 & 46 & 3.1 & 1.0 & 30\% & 54\% & 15\% \\
   \hline
\textbf{Total} & \textbf{14} & \textbf{96} & \textbf{3.2} & \textbf{1.6} & \textbf{40\%} & \textbf{44\%} & \textbf{16\%}
	\end{tabular}

  \vspace{0.05in}

  \caption[Participant interaction with synthesis]{User study participant interaction with synthesis, reporting the number of exercises using synthesis, the number of invocations of the synthesizer, the mean number of assertions and holes when invoked, and the usefulness of results: the percentage of invocations in which the synthesizer returned no result (timeout or crash), returned a result that was completely rejected by the user, and returned a result that was at least partially accepted by the user.}
  \label{table:participantSynthesis}
\end{table}

Participants were familiar with writing assertions. In the first session, the facilitator only introduced participants to providing example function calls, not asserting on their results. Despite this, unprompted, both participants wanted to make assertions once they had an example to work with. When assertions were formally introduced, participants were generally comfortable providing examples, although P1 would occasionally write assertions in a polymorphic form, \eg{} \verb+foldl f acc [] = acc+, which would insert new blank bindings for \verb+f+ and \verb+acc+ on the canvas and P1 would have to recover from the mistake. Even so, P1 appreciated that \ms{} encouraged them to write in a test-driven development (TDD) style, and suspected it prevented them from making simple errors. When asked if they had trouble writing assertions, P2 responded, ``I had trouble \emph{not} making assertions,'' because P2 enjoyed toying with the synthesizer, but P2 did observe that constructing trees was a little tricky. \ms{} currently only beautifies tree values, not tree expressions.
Overall, when we asked how laborious it was to create examples on a scale of 1 to 10, P1 and P2 responded with 2 and 4, respectively. Providing assertions was not a bottleneck.

The facilitator introduced synthesis to the participants with the list \verb+length+ example, which left a positive first impression on the participants.
Synthesis was somewhat less helpful after the \verb+length+ example. Overall, the participants invoked the synthesizer a total of 96 times (6.9 times per exercise), with only 16\% of those invocations returning a result that the user partially or fully accepted (\autoref{table:participantSynthesis}). Despite the low success rate, participants appreciated the synthesizer enough when it succeeded that they were not overly bothered when it did not, and were therefore comfortable invoking synthesis many times.

Before the addition of the ``Accept / Reject'' buttons in the third session there was also no feedback in \ms{} that clearly indicated what had changed—P1 admitted to looking at their Vim window to ascertain what the synthesizer produced. The addition of the ``Accept / Reject'' interface was appreciated by participants and P1 noted they did keep their focus more on the \ms{} window.

Overall, the facilitator's impression was that the participants were comfortable trying to use synthesis, but did not necessarily obtain mastery of it, in part because synthesis is opaque. P1 noted, ``It is really hard to know whether synthesis is failing because I have posed the problem in an incorrect way or synthesis is failing because I haven’t given it a lot of information. But the process of trying to give it more information is very illuminating in terms of whether my conception of the problem is wrong.'' P2 initially felt that working with the synthesizer was unfamiliar but remained intrigued by its potential, saying, ``It was kind of awkward at first. It sort of seemed like a cool trick but there were parts where it would actually complete the program which was kind of nice even though it was not like a very trivial program. That’s a neat feature.'' These experiences suggest that synthesis in this setting is a viable workflow, despite its initial unfamiliarity.

\subparagraph*{RQ4. How do users interact with their text editor?}
Participants were allowed to use their text editor, but the heavy focus on learning \ms{} meant that they only did so only as a last resort. P1 estimated they spent about 40\% of their time looking at Vim when trying to figure out what was going on, but, by the end of the third session, only felt the need to edit in Vim on particularly tricky errors. P2 also felt more comfortable in Vim, ``When I was really stuck, I felt self-conscious and I was like, `Alright I'll just figure this out in Vim quickly.' It's faster, probably, I’ve got years of experience doing that.''

Part of the promise of bimodal editing is that one \emph{can} do this! Even so, \ms{} may be over-reliant on only using shapes and colors to differentiate different kinds of elements, which may have driven the participants to look at their Vim window to understand what was happening instead of relying solely on the \ms{} display.

\subparagraph*{SQ1. What are the pain points? How might the system be improved?}
Participants had trouble keeping track of all the elements on the display. \ms{} relies on colors and shapes to distinguish the multitude of different UI elements: expressions, values, function parameters, assertions, expected values, return expressions, patterns, let-bindings (TVs), and different (sub)canvases that hold let-bindings. Both participants expressed a desire for more explicit labeling of what all these different elements were. After the first session, we added labels on the subcanvases (``Top level'', ``Bindings inside function'', ``Return expression(s) and value(s)'') which P1 expressed appreciation for. We  hoped those would obviate the need for more labeling, but at the end of the final session participants still desired clearer markings.

\subparagraph*{SQ2. Are participants comfortable enough to complete a task without guidance?}
After each exercise we asked participants if they felt comfortable completing the next task without assistance from the facilitator. By the end of the final session P2 was comfortable with minimal assistance, whereas P1 still felt the need for help. Although P1 understood the tool well, they still stumbled over different small issues such as UI corner cases and accidentally trying to edit a parent expression in the subexpression editor (discussed in SQ3 below).

\subparagraph*{SQ3. What else can we learn using the lenses of Cognitive Dimensions of Notations?}
This framework~\cite{CognitiveDimensions} comprises thirteen lenses for qualitatively assessing design trade-offs. Below, we report a subset of our observations from considering these lenses.

\emph{Diffuseness (How noisy is the display?)} \ms{} stores extra information, such as 2D binding coordinates and previously rejected synthesized expressions, as annotations in the OCaml code. Although \ms{} includes a syntax highlighting rule that will gray out these AST annotations, the rule only works in VS Code with the Highlight extension~\cite{HighlightVSCodeExtension} installed. P1 opined that, ``All the annotations do make it less attractive to try to do stuff in Vim.'' Rejected expressions were particularly confusing because they appeared in their entirety in the code, albeit wrapped with \verb+[@not ... ]+. Participants would sometimes read these large expressions thinking it was the code they were writing. After the study, we modified \ms{} to instead store a short hash of the rejected expression.

\emph{Secondary Notation (Is there non-semantic notation to convey extra meaning?)} Currently, \ms{} does not support comments. P1 missed having comments, while P2 did not.

\emph{Viscosity (How hard is it to make changes?)} Three main scenarios arose where changes were difficult. First, editing a base case requires that some execution hits the base case, otherwise the base case can never be focused; this was occasionally a hindrance and might be addressed either by adding a ``phantom call frame'' that focuses the case without a concrete execution or by automatically synthesizing an example that hits the case. Second, once an expression was in the program, it was hard to wrap the existing expression with some new expression; it would be better if there were a mechanism to indicate whether a new drag-and-dropped expression should replace or wrap the old. Finally, although subexpressions can be text-edited by double-clicking them on the display, only that subexpression is opened for editing. Sometimes participants (and the first author) would start editing a subexpression but realize they needed to edit a parent instead. We have since changed \ms{} to open the entire parent for editing but with the clicked subexpression initially selected.

\emph{Visibility (Is everything needed visible? Can items be juxtaposed?)} Element positioning in \ms{} proved tricky, because elements will change size based on the size of the values in the TVs—multiple large trees in the function IO grid, for example, can make a function take up the whole window. Participants did have to move assertions around. P2 used a large screen and expected their functions to grow rightward: P2 would position assertions far to the right of their nascent function. P2 also expressed the desire for snap-to-grid so they could align their TVs perfectly. P1 used a smaller screen which may have caused trouble: at one point P1 was trying to debug and realized after-the-fact that they had scrolled the IO grid offscreen—had it been onscreen and they looked at it, they might have found their mistake quicker. One possible mitigation is to scale down large values.

\section{Related Work}
\label{sec:relatedWork}

Several systems share our goal to center live program values in the programming workflow.

\subparagraph*{Programming by Demonstration (PBD)}
In this interaction paradigm, the user demonstrates an algorithm step-by-step, resulting in a program. The first PBD system, Pygmalion~\cite{Pygmalion}, targeted generic programming and, like \ms{}, displayed the live values in scope as the object of user actions.
For example, a function call with missing arguments was represented as an icon on the canvas.
When all arguments were supplied, the icon was replaced with a display of its result value.
To use that result value, the user dragged the value to where they wanted to use it.
Recursion was supported.
Although the 2D canvas was non-linear, Pygmalion treated the program as an imperative, step-by-step movie over time and did not offer a corresponding always-editable text representation.

Like Pygmalion, Pictorial Transformations (PT)~\cite{PT} also offered  program construction via step-by-step manipulation of live program values. PT allowed the user to customize visualizations, and was generally more expressive than Pygmalion, supporting more complicated algorithms including those involving lists. Later PBD systems were usually more domain-specific~\cite{WWID, YWIMC}, although ALVIS Live~\cite{ALVISLive} targeted iterative array algorithms by demonstration, notably representing the resulting program in editable text. Unlike \ms{}, ALVIS Live generated imperative code and could not offer non-linear editing—UI buttons were needed to allow users to move backwards and forwards in the timeline.

Some empirical evidence of benefits from a value-centric workflow was provided by the Pursuit PBD system for shell scripting~\cite{Pursuit}. In that work, a comic-strip style representation of a program—with before and after values in the frames of a comic-strip—enabled users to more accurately generate programs compared to a more textual representation. On the other hand, when Frison~\cite{AlgoTouchVsPythonTutor} compared student performance between Python Tutor~\cite{PythonTutor}, providing editable code plus live output, and AlgoTouch~\cite{AlgoTouch}, providing non-editable code plus PBD on values, they found students performed comparably in either environment.
An analogous comparison in ALVIS Live also found similar overall student performance when using text or PBD~\cite{ALVISLiveDMStudy}.
These results can be interpreted either way: pessimistically, that value-centric manipulation is not clearly better; or optimistically, that despite non-editable code, value-centric editing performs as well as ordinary programming.
Even so, a PBD environment may aid in avoiding initial fumbling with syntax and in discovering what a tool can do:
Hundhausen et al.~\cite{ALVISLiveDMStudy} found that on the first task, users in the PBD condition worked faster, more accurately, and spent less time consulting documentation.

\subparagraph*{Malleable Live Objects}
In the object-oriented paradigm, the Self~\cite{Self} language and environment displayed live objects graphically, allowing messages to be sent via direct manipulation (demonstrated in video form in \cite{SelfTheMovie}). Although value-centric, the interactions provided by Self and related systems, \eg{} the Morphic UI framework~\cite{Morphic}, differ from all other systems discussed here in that manipulations in Self-like systems modify \emph{state}, not the algorithm.

Like Self, the ``Direct Programming'' prototype~\cite{DirectProgramming} by Edwards allows users to directly invoke actions on displayed values, but, unlike Self, reified these actions in a script, blurring the line between running a program and modifying it. Also blurring the line between runtime interaction and coding, Boxer~\cite{Boxer} was a non-linear programming environment displaying nested boxes on a 2D canvas. A box could contain a comment, code, a value, or serve as a graphics buffer for drawing. Boxes can be edited via code or by user interaction, enabling a workflow that mixes program runtime interaction with program creation. Boxer aimed for its interface to be an approachable computational medium, resulting in design choices that differ from \ms{}. Boxer is not bimodal—the displayed boxes are the program—and state and code are mixed. Also, box results are not automatically rendered. Code boxes must be manually invoked and must write their results to another box, but Boxer includes mechanisms for configuring keys or mouse buttons to trigger particular boxes.

\subparagraph*{Live Nodes-and-Wires}
In 2D nodes-and-wires programming~\cite{TheOnLineGraphicalSpecificationOfComputerProcedures}, nodes usually represent transformations (expressions) and the wires represent dataflow (values).
Consequently, nodes-and-wires environments do not necessarily display live values, although some systems do output live values below the nodes (\eg{} natto.dev~\cite{NattoDev}). Among these environments, Enso~\cite{Enso}, formerly known as Luna, is also bimodal like \ms{}, offering both textual and graphical representations for editing the program.

PANE~\cite{PANE} flips the usual node-and-wires paradigm, instead using example values for nodes and locates transformations (expressions) on wires, placing values more at the center of attention compared to its peers. Example values can be clicked to invoke operations on them. PANE does not, however, maintain an editable text representation of the program.

\subparagraph*{Live Programming}
Like \ms{}, traditional live programming research seeks to augment ordinary, text-based coding with display of live program values, although the displayed values are read-only. There are a growing number of such systems. Python Tutor~\cite{PythonTutor} is a popular teaching tool for visualizing program state in Python and other languages. Bret Victor's Inventing on Principle presentation~\cite{InventingOnPrinciple} demonstrated several live programming environments and served as inspiration for later work~\cite{Seymour,ProjectionBoxes}. Edwards~\cite{ExampleCentricProgramming} showed how examples can be incorporated into the IDE for live execution, and Babylonian-style Programming~\cite{BabylonianStyleProgramming} explored how to better manage multiple examples—individual examples could be switched on and off, an interaction we could adopt in \ms{} to selectively reduce the number of values shown in the function IO grids.

\subparagraph*{In-Editor PBE/PBD}
Like our programming by example (PBE) synthesizer, recent work explores PBE and PBD interactions in textual environments.
Several systems generate code within a computational notebook via manipulation of visualized values. Wrex~\cite{Wrex} adapts the FlashFill~\cite{FlashFill} PBE workflow to Pandas dataframes in Jupyter notebooks—after demonstrating examples of a desired data transformation in a dataframe spreadsheet view, Wrex outputs readable Python code. Similarly, the PBD systems B2~\cite{B2}, mage~\cite{mage}, and Mito~\cite{Mito} transform step-by-step interactions on displayed notebook values into Python code. For a Haskell notebook environment, Vital~\cite{Vital1,Vital2} offers copy-paste operations on visualized algebraic data type (ADT) values, which are realized by changing the textual code in the appropriate cell. Like \ms{}, graphical interactions in Vital can extract subvalues via pattern matching, although Vital's workflow focuses on modifying single values in place rather than building up computations like in \ms{}. While these notebook systems provide some manipulation of intermediate values, none offer fine-grained non-linearity.

For more ordinary IDE settings, CodeHint~\cite{CodeHint}, SnipPy~\cite{SnipPy},  and JDial~\cite{JDial} provide synthesis interactions in the live context of the user's incomplete code. With CodeHint, users set a breakpoint in their Java program and describe a property about a desired value—CodeHint enumerates method calls in the execution environment at the breakpoint to find a satisfying expression. Like \ms{}, CodeHint uses a statistics model to rank results. Notably, users with CodeHint were significantly faster and more successful at completing given tasks than users without.
For Python, SnipPy~\cite{SnipPy} adapts the Projection Boxes tabular display of runtime values~\cite{ProjectionBoxes} to perform PBE in the context of live Python values. The authors validated that users successfully used the synthesizer to complete portions of given tasks.
JDial~\cite{JDial} records variable values during execution of an imperative Java program and allows the programmer to directly change incorrect values in the trace. The program is repaired to match the corrections via sketch-based synthesis~\cite{SketchingThesis}. JDial, however, is limited to small program repairs and does not offer program construction features.

\subparagraph*{Bidirectional, Bimodal Programming}
Some systems represent programs as ordinary text, but also allow direct manipulation on program \emph{outputs} to be back-propagated to change the original code. Usually, these changes are ``small'' changes to literals in the program—such as numbers~\cite{sns-pldi,CarbideAlpha,sns-oopsla,CapStudio},  strings~\cite{AutomatingPresentationChangesInDynamicWebApplicationsViaCollaborativeHybridAnalysis,LiveProgrammingByExampleUsingDirectManipulationForLiveProgramSynthesis,CarbideAlpha,sns-oopsla}, or lists~\cite{sns-oopsla}.
More full-featured program construction via output manipulations is available in a few systems for programs that output graphics, such as APX~\cite{APX}, Transmorphic~\cite{Transmorphic}, and Sketch-n-Sketch~\cite{sns-uist-2019}.

Although \ms{} also centers values as subjects for manipulation, we do not yet apply bidirectional techniques to deeply back-propagate a change on a value—direct changes on a value are only allowed when it was introduced as a literal in the immediately associated expression. An earlier version of \ms{} did support limited back-propagation, which we disabled because it caused trouble in the user study: manipulating a value would inconspicuously change a literal in a very different part of the program. Determining an understandable meaning of such direct changes on a value remains an avenue for future work.
While the bidirectional programming community offers the ``least change'' principle~\cite{Meertens98}, that minimal changes should be performed to maintain a constraint, in the context of a full program a change may cause confusion not because of its magnitude but because the item changed is far from the user's focus. Revealing that far-away code by popping open a ``bubble''~\cite{CodeBubbles1,CodeBubbles2} or ``portal''~\cite{CodePortals} may be one way to help make the change understandable.

\section{Future Work and Conclusion}
\label{sec:last}

\subparagraph*{Scaling Up}
So far, \ms{} has been applied only to small, side-effect-free programs. Running larger programs requires managing traces efficiently, and handling practical programs requires managing side effects such as input/output and mutation.

Tracing has the potential to consume a considerable amount of memory: every execution step produces a new portion of the trace. Currently, traces are all stored in RAM. Future tracing could instead write to persistent storage.
Furthermore, for a large program, only a small portion of the trace will likely be needed at a time. Tracing can be skipped for code outside the region of interest. When needed, missing portions of the trace could be rebuilt on demand via program replay, which can be accomplished efficiently by periodically dumping the program state during the initial execution and replaying from a checkpoint as needed~\cite{EfficientAlgorithmsForBidirectionalDebugging}.

Support for side effects requires both changes to the UI in addition to technical engineering. Side effects are necessarily linear, whereas \ms{}'s UI is currently based around a free-form, unordered 2D canvas. One UI possibility is to have each imperative statement spatially divide its (sub)canvas into ``before'' and ``after''---TVs spatially above the imperative statement are executed before it, and TVs below are executed afterwards.
For the implementation, careful interception and logging of system calls can record all program I/O and enable deterministic replay, even for large programs~\cite{rr}. In \ms{}, recording and replaying side effects could be handled in the interpreter rather than at the system call level.

\subparagraph*{Value-Oriented Thinking}
We hypothesize that expression-oriented and value-oriented modes of thinking are distinct states of mind, and experienced programmers tend towards the former. An intriguing possibility for future work is to experimentally validate that expression-oriented and value-oriented thinking are actually modes—\ie{} the activity of considering values discourages considering expressions, and vice versa.
More immediately, there are possible changes to \ms{} that might encourage more value-focused interaction.

One experiment is to change the display of variable uses so that, instead of the name of the variable, its current value is shown instead, with the name as a tooltip or subscript. This change might nudge users out of the expression-oriented mode of thinking back towards value-oriented thinking.
An intriguing corollary experiment was requested by P1. To keep track of their provenance, P1 wanted values to be drawn with unique colors all the time, rather than only when autocomplete menus were open. Another possibility is, when the cursor is over a variable usage, to highlight the TV where the variable is defined. We would like to explore these display choices, as well as other opportunities for ``linked'' visualizations~\cite{Roly2022}.

Finally, while dragging items onto an expression is quite useful, dragging items onto values is currently less so. When working through the examples, the implementer dragged some item onto a value on only 4 occasions, compared to dragging onto an expression 209 times. In the future, dragging a value to a value might open a menu of possible ways to combine the values. Ideally, programmers should be able to customize the available actions, as in Vital~\cite{Vital2} which includes an API for this purpose.

\ms{} currently focuses on interactions on relatively small values. Larger data sets might be more more conveniently displayed and manipulated in a spreadsheet-style view---Flowsheets~\cite{Flowsheets} demonstrates one such approach, albeit without program synthesis.

\subparagraph*{Conclusion}
How close is \ms{} to achieving its goals of providing a value-centric, non-linear programming environment? Based on the examples we implemented and feedback from our study participants, \ms{} largely succeeded at providing useful live values. The non-linear features functioned moderately well—users rarely had to think about binding order—but \ms{} was not immediately learnable and would benefit from more explicit labeling of the various kinds of elements on the canvas.

Overall, building on the insight from Eros~\cite{Eros} that non-linearity complements functional programming, \ms{} shows that non-linearity can be maintained even when the program is ordinary code. Our emphasis on textual code has resulted in \ms{} currently being somewhat more expression-centric than Eros.
As described above, there are many possible ways \ms{} might become more value-centric, to further our vision to make programming feel like a tangible process of molding and forming.

\subparagraph*{Acknowledgements}
We extend our gratitude to Justin Lubin for advice (so far ignored) about caching during synthesis,
Byron Zhang for feedback on the presentation,
Matt Teichman for providing hardware,
Kartik Singhal for technical support, and
the user study participants for their great patience with bugs and for the invaluable feedback they provided.
This work was supported by U.S. NSF grant \emph{Direct Manipulation Programming Systems} (CCF-1651794).

\bibliography{references}

\vfill
\pagebreak

\appendix

\section{Subvisualizations}

\begin{figure}[h]
  \centering
  \begin{tabular}{ccc}
  \includegraphics[height=1.5in]{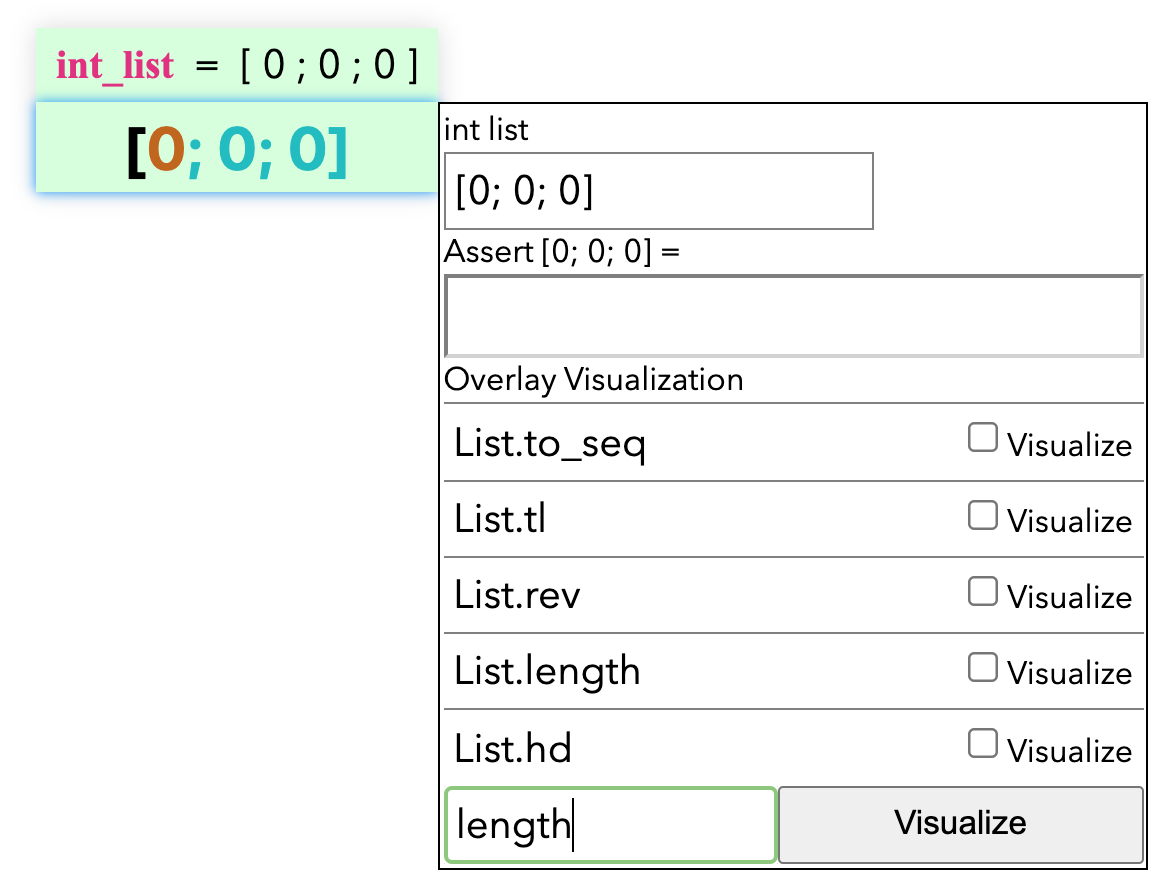} &
  \includegraphics[height=1.2in]{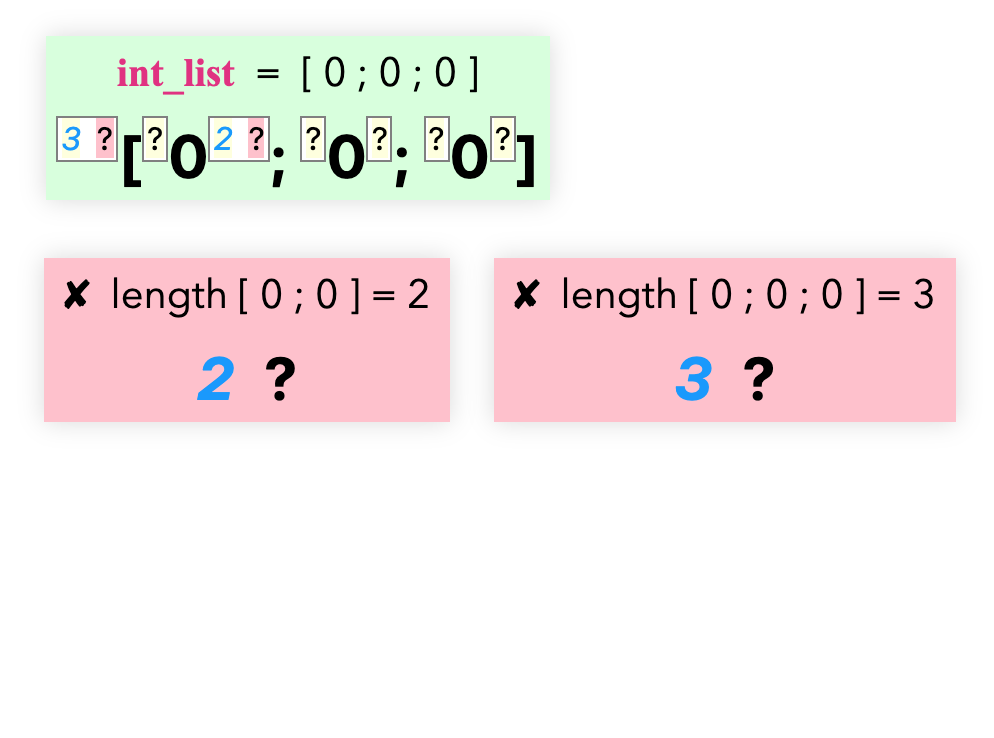} &
  \includegraphics[height=1.2in]{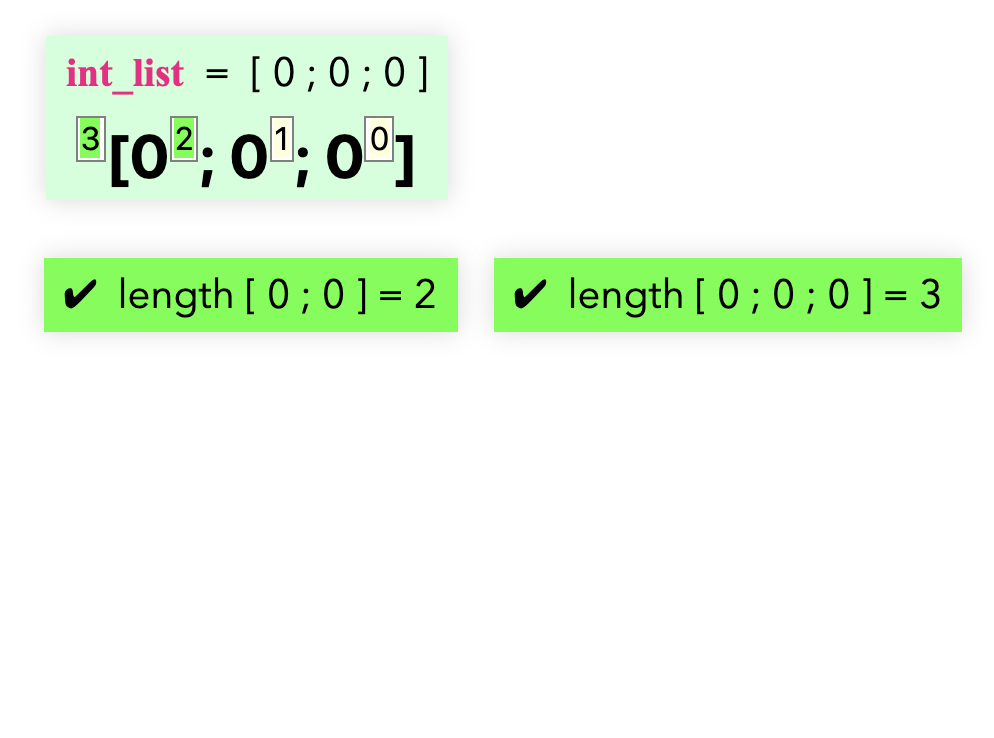} \\
  (a) & (b) & (c)
  \end{tabular}

  \caption[Subvisualizations]{(a)~Adding a subvisualization, with assertions (b)~before and (c)~after satisfaction.}
  \label{fig:subvisualizations}
\end{figure}

As a simple way to customize the visualization of values,
\ms{} also offers users the ability to visualize the result of a function call on all subvalues of a displayed value—for example, the result of calling \verb+length+ on each sublist of \verb+[0; 0; 0]+ can be displayed as superscripts on the sublists. Assertions can also be specified on these visualized results, leading to the following alternative workflow for the \nameref{sec:overview}:

As in the original workflow, Baklava first inserts \verb+[0; 0; 0]+ on her blank canvas. Afterward, she instead clicks the \verb+[0; 0; 0]+ value to select it. A floating inspector window offers various type-compatible functions in scope to visualize atop \verb+[0; 0; 0]+. Baklava forgoes these suggestions and navigates to the textbox that allows her to provide a custom subvisualization (\autoref{fig:subvisualizations}a). She types ``length'' and hits Enter. The length function skeleton is automatically created, and each subvalue of \verb+[0; 0; 0]+ now displays a superscript \verb+?+, the return result of the unfinished \verb+length+ function when applied to that subvalue. Baklava double-clicks the superscript for the whole \verb+[0; 0; 0]+ value, which opens a textbox that allows her to assert on \verb+length [0; 0; 0]+. She types ``3'', hits Enter, and the appropriate assertion is created at the top-level.

These subvisualizations allow users to quickly specify multiple assertions without manually creating new example values. To assert on a list of length 2, Baklava double-clicks the superscript for the tail sublist and types ``2'' (\autoref{fig:subvisualizations}b). With these two assertions, the synthesizer finds the intended result in one try (\autoref{fig:subvisualizations}c shows the satisfied assertions).

\section{Handling Non-Linearity}
\label{sec:handlingNonlinearity}

To provide a non-linear editing experience, \ms{} automatically reorders bindings based on names, normalizes \verb+match+ statements, and inserts new bindings for undefined variables. These processes are detailed below.

\subparagraph*{Reordering Bindings}

\ms{} assumes all names at each ``indentation scope'' in the code are unique and unordered. For example, in \autoref{fig:reorderingExample}a, the top-level names \verb+a+, \verb+b+, and \verb+c+ are in the same indentation scope, and the names \verb+x+ and \verb+a+ within the definition of \verb+c+ are in a indentation scope.

Names must be unique within the indentation scope so that the names, not the order, indicates the dependencies. After each action, \ms{} reorders bindings to reify the dependences as valid OCaml with proper lexical scoping, as demonstrated in \autoref{fig:reorderingExample}b\&c. This rearrangement algorithm is described below.
\\

\noindent
\textit{Starting from the first binding in the ``indentation scope''...}
\begin{enumerate}
  \item The right-hand-side (RHS) of the binding is recursively rearranged according to this algorithm. After rearrangement, any remaining free variables in the RHS indicate names not defined within, but still needed by, the RHS. Let $\mathcal{N}$ be this set of needed names.
  \item If any names in $\mathcal{N}$ match a variable defined in this binding's LHS pattern, a \verb+rec+ flag is added to the binding to marked it as recursive. These LHS names are removed from $\mathcal{N}$.
  \item The LHSes of all later bindings at this binding's indentation scope are examined to see if they define any names in $\mathcal{N}$.
  \item If one such binding is found, it is moved above the current binding. The algorithm loops back to Step 1 with the newly moved binding as the binding of interest (so that any of its dependencies will be recursively moved as well).
  \item If no such binding is found, the algorithm proceeds to the next binding in the indentation scope and repeats from Step 1.
\end{enumerate}

Single recursion is handled by Step 3, but mutual recursion is not (yet) supported. A cycle check is used in Step 4 to prevent the algorithm from diverging if two bindings are mutually dependent on each other, but the bindings are not gathered into a mutually recursive group.

\subparagraph*{Handling Case Splits}

Recall that users can grab any displayed subvalue and drag it into their program to induce a pattern match. Internally, the process works as follows. Whenever a user hovers their mouse over a value, a tooltip appears previewing the expression that will be inserted. For subvalues, the expression is an incomplete pattern match, such as \texttt{match list with | hd::tail -> tail}. Deeper extractions are also possible, \eg{} \texttt{match (match list with | hd::tail -> tail) with | hd2::tail2 -> tail2}, but not often useful.

When the user drops the subvalue into their program, the expression is initially (internally) inserted as is, \eg{} \texttt{let tail2 = match list with | hd::tail -> tail in ...}. A series of program transforms then rearranges \verb+match+ statements as follows:

\begin{enumerate}
  \item All let-bindings at the beginning of the function are pushed down and duplicated into each pre-existing case branch. They may be pulled back out at the end of the process below. This push-down has the effect that all newly inserted \verb+match+ statements are now children of any prior \verb+match+ statements already inserted.
  \item If the user dragged a subvalue that has already been extracted (or has extracted a deeper subvalue and one of its parents has already been extracted), we do not want to insert superfluous pattern matches. We want to reuse the case splits that already exist. Relying on the prior step that ensured all bindings are now in scope of all variables introduced in cases, a static analysis pass simplifies nested case splits on the same variable: if a \verb+match+ on \verb+list+ already exists, then the copy of \texttt{let tail2 = match list with | hd::tail -> tail in ...} that was pushed into the pre-existing cons case will be simplified to \texttt{let tail2 = tail in ...} and the copy pushed into the pre-existing empty-list case will be simplified to \texttt{let tail2 = match list with in ...}, \ie{} a match with no cases, which marks the binding for removal.
  \item Each let-binding that has such an empty \verb+match+ anywhere in its left-hand-side is removed.
  \item Any surviving \verb+match+ statement is not redundant, but still in a non-idiomatic position. A series of local rewrite rules floats the \verb+match+ statements up to the outermost level of the function, \eg{} \texttt{f (match list with hd::tail -> tail)} becomes \texttt{match list with hd::tail -> f tail}, etc.
  \item All let-bindings, previously floated down into all case branches, are now floated back up as far as possible to the top-level of the function: a binding is pulled up outside of \verb+match+ branches when both (a) the same binding exists in all branches, \ie{} it was valid in all branches and not removed, and (b) in all branches, the binding is not dependent (or transitively dependent) on any variables introduced for the case branch.
  \item Newly inserted \verb+match+es are now at the top-level, but may still be missing cases. Incomplete pattern matches are filled in with the missing branches, with a hole expression \verb+(??)+ in each new case.
  \item Bindings that are only simple renamings, such as \verb+let tail2 = tail in ...+, are removed—these happen when the user performs an extraction of a subvalue that was already previously extracted.
\end{enumerate}

The above algorithm produces idiomatic \verb+match+ statements, with the \verb+match+ wrapped in all the let-bindings that are not dependent on variables introduced in the branches. For functions with a single \verb+match+, the above algorithm performed well in our evaluation—it was never a source of trouble. For nested matches with independent scrutinees, a current limitation of \ms{} is there is no refactoring to flip the nesting order.

\subparagraph*{New Variable Insertion}

After reordering bindings and handling case splits, a new let-binding is inserted for any remaining undefined variables. Each new variable is either bound to hole or, if the variable is used as a function in an application, bound to a new function skeleton with the appropriate number of parameters.

\section{Synthesis PCFG}

\begin{figure}[bt]

\newcommand{\termOr}{\ |\ }
\newcommand{\kw}[1]{\text{\textbf{#1}}}
\definecolor{probColor}{rgb}{0.07,0.43,1.0}
\newcommand{\prob}[1]{\textcolor{probColor}{\ensuremath{^{\scriptstyle#1}}}}
\newcommand{\chr}[1]{\texttt{'#1'}}
\newcommand{\str}[1]{\texttt{"#1"}}
\newcommand{\bkslsh}{\char`\\}

\newcommand{\additionalGrammerLine}[1]{\\ & $\termOr$ & $#1$}
\newcommand{\grammerLine}[3]{\textbf{#1}\hspace{0.03in} $#2$ & $::=$ & $#3$}
\newcommand{\grammerKindSep}{\vspace{0.03in}\\}

\newcommand{\longMetaVar}[1]
  {\ensuremath{\mathit{#1}}}

\centering
{\renewcommand{\arraystretch}{0.8} 
\begin{tabular}{rll}

\grammerLine{Expressions}{e}{\prob{52\%}x}
\additionalGrammerLine{\prob{20\%}e_1\ \overline{e_i}}
\additionalGrammerLine{\prob{10\%}\kw{fun}\ x \rightarrow e}
\additionalGrammerLine{\prob{8.1\%}\longMetaVar{ctor}}
\additionalGrammerLine{\prob{6.6\%}c}
\additionalGrammerLine{\prob{1.9\%}\kw{match}\ e_1\ \kw{with}\ \overline{C ... \rightarrow\phantom{t}e_i}}
\additionalGrammerLine{\prob{1.3\%}\kw{if}\ e_1\ \kw{then}\ e_2\ \kw{else}\ e_3}

\\
\grammerKindSep{}
\grammerLine{Names}{x}{\prob{73\%}\longMetaVar{localName} \termOr \prob{27\%}\longMetaVar{pervasivesName}}

\grammerKindSep{}
\grammerLine{Local Names}{\longMetaVar{localName}}{\prob{31\%}\longMetaVar{MostRecentlyIntroduced}}
\additionalGrammerLine{\prob{20\%}\longMetaVar{2ndMostRecentlyIntroduced}}
\additionalGrammerLine{\prob{11\%}\longMetaVar{3rdMostRecentlyIntroduced}}
\additionalGrammerLine{\text{...etc...}}

\grammerKindSep{}
\grammerLine{Pervasives Names}{\longMetaVar{pervasivesName}}{\prob{4.0\%}\texttt{(+)} \termOr \prob{2.9\%}\texttt{(=)} \termOr \prob{1.9\%}\texttt{(-)} \termOr \prob{1.4\%}\texttt{(\&\&)} }
\additionalGrammerLine{ \prob{1.3\%}\texttt{(||)}\prob{1.3\%}\texttt{(\^{})} \termOr \prob{1.2\%}\texttt{not} \termOr  \prob{1.2\%}\texttt{(<)} \termOr \prob{0.90\%}\texttt{(<>)} }
\additionalGrammerLine{\prob{0.63\%}\texttt{(\textrm{@})} \termOr \text{...other non-imperative names...}}
\\
\grammerKindSep{}
\grammerLine{Constructors}{\longMetaVar{ctor}}{\prob{54\%}\longMetaVar{pervasivesCtor} \termOr \prob{46\%}\longMetaVar{userCtor}}

\grammerKindSep{}
\grammerLine{Pervasives Ctors}{\longMetaVar{pervasivesCtor}}{e_1 \prob{24\%}\hspace{-1pt}\text{::} e_2 \termOr \prob{20\%}\texttt{[]} \termOr \prob{15\%}\texttt{()} \termOr \prob{13\%}\texttt{false}}
\additionalGrammerLine{ \prob{8.2\%}\texttt{true} \termOr \prob{7.2\%}\texttt{None} \termOr \prob{6.7\%}\texttt{Some}\ e \termOr \text{...etc...}}
\grammerKindSep{}
\grammerLine{User Ctors}{\longMetaVar{userCtor}}{\text{...ctors defined in file (uniform probability)...}}

\\
\grammerKindSep{}
\grammerLine{Constants}{c}{\prob{52\%}\longMetaVar{int} \termOr \prob{45\%}\longMetaVar{str} \termOr \prob{2.2\%}\longMetaVar{char} \termOr \prob{0.43\%}\longMetaVar{float}}

\grammerKindSep{}
\grammerLine{Int Literals}{\longMetaVar{int}}{\prob{37\%}0 \termOr \prob{26\%}1 \termOr \prob{10\%}2 \termOr \prob{3.4\%}3}
\additionalGrammerLine{\prob{2.7\%}4 \termOr \prob{1.6\%}8 \termOr \prob{1.5\%}5 \termOr \prob{1.1\%}$-$1}
\grammerKindSep{}
\grammerLine{String Literals}{\longMetaVar{str}}{\prob{2.9\%}\str{} \termOr \prob{0.62\%}\str{.} \termOr \prob{0.59\%}\str{)} \termOr \prob{0.37\%}\str{ }}
\additionalGrammerLine{\prob{0.34\%}\str{(} \termOr \text{...other 1-char strs from corpus...}}
\grammerKindSep{}
\grammerLine{Char Literals}{\longMetaVar{char}}{\prob{9.1\%}\chr{\bkslsh{}n} \termOr \prob{8.4\%}\chr{ } \termOr \prob{6.2\%}\chr{\bkslsh{}\bkslsh{}} \termOr \prob{4.2\%}\chr{0}}
\additionalGrammerLine{\prob{4.2\%}\chr{-} \termOr \text{...other chars from corpus...}}
\grammerKindSep{}
\grammerLine{Float Literals}{\longMetaVar{float}}{\prob{25\%}0.0 \termOr \prob{20\%}1.0 \termOr \prob{18\%}0.0 \termOr \prob{8.3\%}0.5 \termOr \prob{5.0\%}10.0}

\end{tabular}
}

\caption[Grammar used for the statistics model]{Grammar used for the statistics model, with probabilities associated with productions.}
\label{fig:pcfg}
\end{figure}

Synthesized programs are scored according to the probabilistic context-free grammar (PCFG) given in \autoref{fig:pcfg}. We subdivided several kinds of terms into multiple production rules in order to provide more precise probabilities: constants are divided by the constant type, constructors are classified as either a user constructor (\ie{} defined in the same module or a parent module) or from elsewhere, and, most notably, names are classified as local (\ie{} defined in the same module or a parent module) or from elsewhere. User constructors are assigned equal probability with each other. Local names are denoted by how recently they were introduced into the execution environment (\ie{} how spatially close they are in the code)—nearby names are much more probable than names introduced far away. The probabilities of constant literals and external (\eg{} standard library) names and constructors are derived directly from how often those names and constructors appear in the corpus.

After calculating these probabilities from the corpus, we further reduced the search space for the synthesizer. We unscientifically trimmed down the list of possible constant literals to the 5-10 most common in the corpus for each type. We also limited the initial execution environment to constructors and functions in the globally-imported \verb+Pervasives+ module, removing functions involving imperative features (they are not supported by \ms{}) as well as several floating point primitive operators unimplemented by Camlboot. Finally, we also excluded OCaml's polymorphic \verb+compare+, which, in practice, the synthesizer would often use in surprising ways to produce the numbers $0$ and $1$, \eg{} \verb+compare x x+ evaluates to \verb+0+. For simplicity, we did not renormalize probabilities after the above trimmings to the production rules.

\section{Example Push-Down for Synthesis}

As part of the type-directed refinement process, \textsc{Myth} will push the user's examples to the frontier of synthesis. For example, if the user asserts that a hole should output \verb+0+ when given the input \verb+[]+, \textsc{Myth} will refine the hole to \verb+fun x -> (??)+ and refine the example to note that \verb+(??)+ should resolve to \verb+0+ when \verb+x+ is bound to \verb+[]+. This allows \textsc{Myth} to quickly verify when a hole filling satisfies all given examples.

However, \textsc{Myth}'s implementation of this example refinement machinery requires that the user provide all assertions directly on holes. Users cannot write \verb+assert (length [] = 0)+. Instead, they must write, essentially, \verb+let length = ((??) such that { [] => 0 })+. It would be better if users could invoke synthesis on a partial sketch.

To allow assertions on program sketches rather than only on holes, \textsc{Smyth} uses a technique dubbed ``live unevaluation''~\cite{Smyth} to push top-level assertions down to constraints directly on holes. Pushing down the assertions is not fundamentally required to perform synthesis—a synthesizer may guess terms at the holes and check the top-level assertions (indeed \ms{} does so)—but pushing the assertions down to the holes provides information about the hole. \ms{} adapts an effectively\footnote{Some sketches prevent constraint propagation. For example, if a sketch has a hole in scrutinee position, it is impossible to know which branch to take. The push-down procedure is stuck and any holes in the branches do not receive their constraints. \textsc{Smyth} resolves this scenario by speculatively filling the scrutinee hole while pushing down the constraints. We opt to avoid this extra machinery, resulting in a simpler implementation, but at the cost that we cannot resolve holes in an iterative fashion—satisfying the requirements of one hole before moving on to the next—which would enable faster synthesis.} identical approach to \textsc{Smyth}, pushing examples through the sketch to yield constraints directly on holes. \ms{} uses these hole constraints for two purposes:

\begin{enumerate}
	\item Refining a hole into a function requires knowing that the hole is at arrow type. This can easily be determined if the program is explicitly typed, as required in \textsc{Myth} and \textsc{Smyth}. \ms{}, however, allows untyped sketches which often start at polymorphic type. For example, in an initial sketch \verb+let length = (??)+, the \verb+length+ variable has type \verb+'a+. When \verb+assert (length [] = 0)+ is pushed down to the hole, we know the hole must satisfy the requirement $[] \Rightarrow 0$, \ie{} must be a function that when given \verb+[]+ produces \verb+0+. If all the constraints on a hole are of that form $v_1 \Rightarrow v_2$, then \ms{} will attempt to refine the hole into \verb+fun x -> (??)+.
	\item To speed up synthesis and produce more relevant results, \ms{} tracks whether a generated (sub)term is allowed to be constant or not (\eg{} \ms{} requires that at least one argument in a function call must be non-constant). If multiple different examples reach a hole, \ms{} will not generate a constant term for that hole. Similarly, if a single example reaches a hole, \ms{} will exclude all constants from consideration for that hole, except for the value asserted on the hole.
\end{enumerate}

\end{document}